\documentclass[manuscript]{acmart}
\usepackage[utf8]{inputenc}
\usepackage{balance}
\usepackage{graphicx}
\usepackage{subcaption}
\usepackage[dvipsnames]{xcolor}
\usepackage{multirow}
\usepackage{array}
\usepackage{breqn}
\usepackage{mathtools}
\usepackage{amsmath}
\usepackage[ruled,vlined,linesnumbered]{algorithm2e}
\usepackage{tikz}
\usepackage{pifont}
\usepackage{tikz}
\usetikzlibrary{arrows.meta, positioning, shapes.multipart, shapes.symbols}

\labelformat{subfigure}{\thefigure(#1)}
\newcolumntype{P}[1]{>{\centering\arraybackslash}p{#1}}
\AtBeginDocument{%
  }

\setcopyright{acmlicensed}
\copyrightyear{2018}
\acmYear{2018}
\acmDOI{XXXXXXX.XXXXXXX}
\acmConference[Conference acronym 'XX]{Make sure to enter the correct
  conference title from your rights confirmation email}{June 03--05,
  2018}{Woodstock, NY}
\acmISBN{978-1-4503-XXXX-X/2018/06}

\begin{document}

\title{AutoTour: Automatic Photo Tour Guide with Smartphones and LLMs}

\author{Huatao Xu}
\affiliation{%
 \institution{The Hong Kong University of Science and Technology}
  \country{Hong Kong SAR}
}
\email{huatao@ust.hk}
\orcid{0000-0002-8289-0910}

\author{Zihe Liu}
\affiliation{%
 \institution{The Hong Kong University of Science and Technology}
  \country{Hong Kong SAR}
}
\email{zzengar@connect.ust.hk}
\orcid{}

\author{Zilin Zeng}
\affiliation{%
 \institution{The Hong Kong University of Science and Technology}
  \country{Hong Kong SAR}
}
\email{zzengar@connect.ust.hk}
\orcid{0009-0005-3592-2587}

\author{Baichuan Li}
\affiliation{%
 \institution{The Hong Kong University of Science and Technology}
  \country{Hong Kong SAR}
}
\email{zzengar@connect.ust.hk}
\orcid{}

\author{Mo Li}
\authornote{Mo Li is the Corresponding author.}
\affiliation{%
  \institution{The Hong Kong University of Science and Technology}
  \country{Hong Kong SAR}
}

\email{lim@ust.hk}
\orcid{0000-0002-6047-9709}


\begin{abstract}
We present \textit{AutoTour}, a system that enhances user exploration by automatically generating fine-grained landmark annotations and descriptive narratives for photos captured by users. The key idea of AutoTour is to fuse visual features extracted from photos with nearby geospatial features queried from open matching databases. Unlike existing tour applications that rely on pre-defined content or proprietary datasets, AutoTour leverages open and extensible data sources to provide scalable and context-aware photo-based guidance. To achieve this, we design a training-free pipeline that first extracts and filters relevant geospatial features around the user’s GPS location. It then detects major landmarks in user photos through VLM-based feature detection and projects them into the horizontal spatial plane. A geometric matching algorithm aligns photo features with corresponding geospatial entities based on their estimated distance and direction. The matched features are subsequently grounded and annotated directly on the original photo, accompanied by large language model–generated textual and audio descriptions to provide an informative, tour-like experience. We demonstrate that AutoTour can deliver rich, interpretable annotations for both iconic and lesser-known landmarks, enabling a new form of interactive, context-aware exploration that bridges visual perception and geospatial understanding.

\end{abstract}




\received{20 February 2007}
\received[revised]{12 March 2009}
\received[accepted]{5 June 2009}

\maketitle

\section{Introduction}

Tourism is an essential and irreplaceable aspect of modern life. Traditional tour and travel applications predominantly focus on delivering pre-curated tours, guided narratives, and static map-based exploration. For instance, popular applications such as SmartGuide \cite{smartguide2025}, GetYourGuide \cite{getyourguide2025}, and Viator \cite{viator2025} provide location-based information and audio-guided experiences but lack real-time interactivity and personalized content. Meanwhile, platforms like Google Maps \cite{googlemaps2025} and Mapy.com \cite{mapy2025} offer offline maps and point-of-interest tagging, yet they require extensive navigation and manual interaction with the map interface.

\begin{figure}[t!]
    \centering
    \makebox[\textwidth][c]{%
      \includegraphics[width=0.7\textwidth]{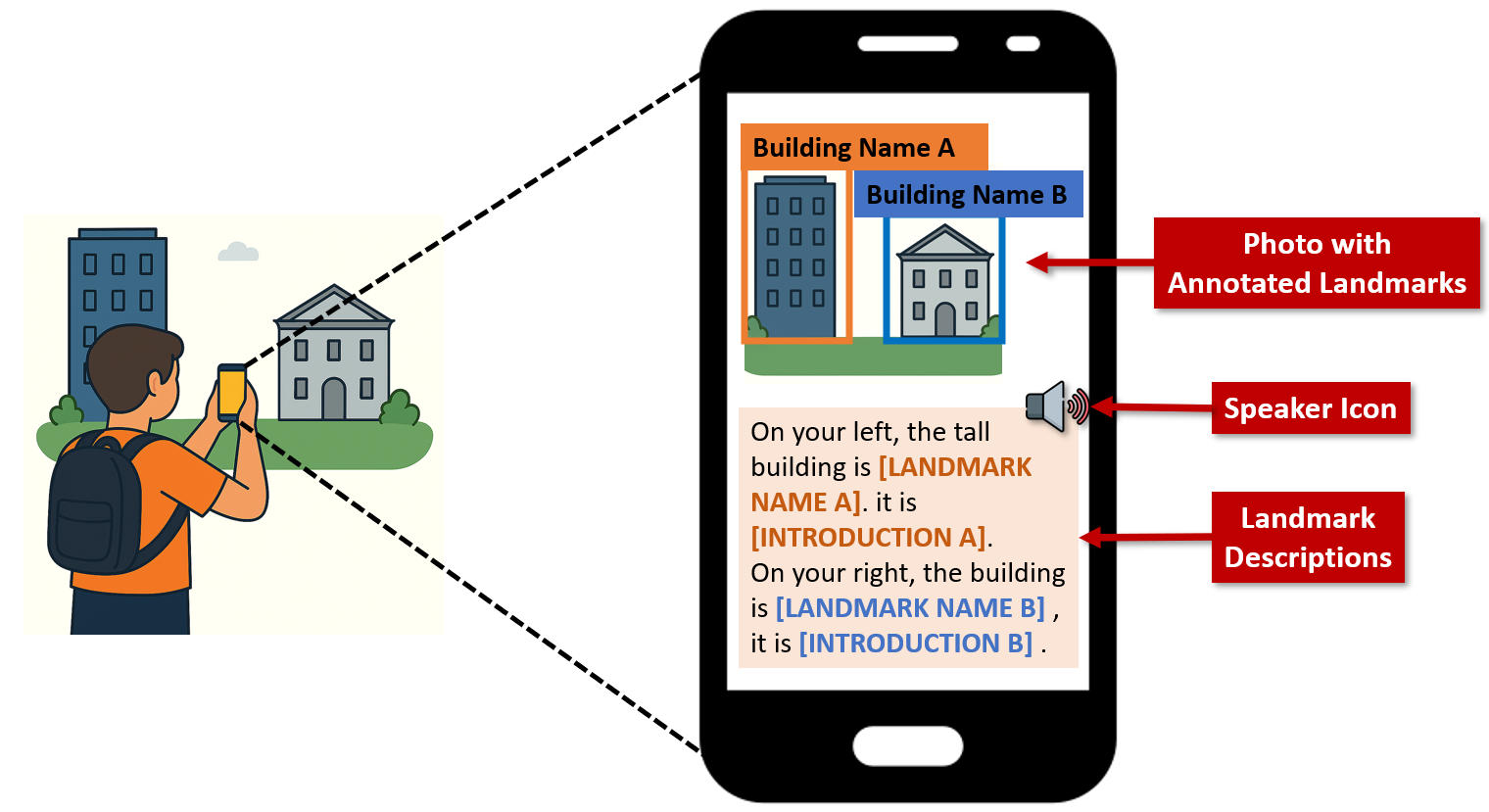}
    }
    \caption{Illustration of a new user interaction approach for touring.}
    \label{fig:scenario}
\end{figure}

In this paper, as illustrated in Fig. \ref{fig:scenario}, we propose a novel interaction way to enhance user exploration through intelligent photo-based guidance. Users simply capture photographs using their smartphones, and the application automatically annotates key landmarks and natural features, such as buildings, lakes, and other landmarks, directly onto the images. In addition, it also presents the descriptions for those landmarks that can be translated to audio for people with needs. Those advancements not only personalize the travel experience but also deepen the user's connection to their surroundings by offering instant, context-aware information.

Although current deep learning models offer foundational capabilities for object recognition and scene understanding, they exhibit notable limitations in the context of personalized and fine-grained landmark detection. Traditional object detection models, such as YOLO \cite{redmon2016you} and Faster R-CNN \cite{ren2015faster}, are proficient at identifying general categories like "person," "car," or "tree," but they lack the semantic granularity required to distinguish specific landmarks or buildings by name. Vision-language models (VLM) like GPT-4o \cite{openai2025gpt4o} and CLIP \cite{radford2021learning} have advanced the field by integrating visual and textual data, enabling them to recognize and describe well-known landmarks such as the Statue of Liberty or the Eiffel Tower. However, these models often struggle with less-famous or region-specific structures due to their reliance on large-scale, general-purpose datasets that may not encompass the diversity of global architecture.

To this end, we propose \textit{AutoTour}, a system that delivers fine-grained landmark annotations and detailed descriptions for photos captured by users. The key idea is to fuse visual information from photos with nearby objects retrieved from geospatial databases. However, to the best of our knowledge, there is currently no global dataset that contains comprehensive landmark names and structures, posing a significant barrier to training deep learning models for this large-scale task. To overcome this challenge, we design a \textit{training-free system} that leverages the strong perception capabilities of existing VLMs and integrates them with open geo-spatial data sources (GDS) that maintain extensive databases of buildings and landmarks worldwide.

Nevertheless, such integration introduces several challenges. Existing GDS platforms, such as Google Maps or AMap, are widely used for localization and navigation but are proprietary, offering only limited APIs that fail to provide access to comprehensive landmark-level information. To this end, we leverage \textbf{OpenStreetMap (OSM)}, an open-source geospatial platform that provides detailed data and supports flexible queries through the Overpass API. However, data retrieved from OSM—typically in JSON or XML format—is comprehensive yet challenging to process effectively. To address this, we design a dedicated preprocessing pipeline that extracts key \textit{geospatial features} (e.g., buildings, roads, parks) from OSM data, filters them based on the user’s GPS location and camera orientation using on-device sensor data (e.g., GPS and IMU), and derives essential attributes such as feature type, distance, and direction to the user.

Another major challenge arises from the discrepancy between spatial perspectives: photographs are captured from a \textit{vertical} viewpoint, whereas geo-spatial data is represented in a \textit{horizontal} plane. To address this issue, \textit{AutoTour} adopts a divide-and-conquer strategy. We simplify the vision task by focusing on the detection of primary buildings or landmarks (termed \textit{photo features} in this paper) and estimating their approximate distance, direction, and type — effectively projecting them from the vertical photo plane into the horizontal geospatial plane. The system then performs geometric overlap matching between photo features and geospatial features based on estimated distance and direction. Finally, matched features are visually annotated on the original photo using vision grounding models~\cite{Qwen2.5-VL,li2021referring}, and large language models generate descriptive narratives, collectively delivering an interactive and informative guided-tour experience.





We implement a functional mobile application that allows users to freely take photographs and display descriptions and annotated features displayed on the map. To evaluate our system, we collect a dataset comprising 125 photos from multiple cities in China and the United States. Comprehensive user studies are conducted, where volunteers rate the quality of the generated annotations and descriptions. Our system achieves an average rating of 3.6 out of 4, demonstrating its effectiveness and user satisfaction. In summary, this work makes the following key contributions:
\begin{itemize}
    \item We explore a practical and user-friendly human-computer interaction approach to help users interpret their surroundings through photography. 
    \item We propose a novel framework that aligns geo-spatial and visual features by leveraging the powerful, training-free capabilities of VLMs and a geometric overlap matching algorithm.
    \item We develop a working prototype of AutoTour and validate it through extensive experiments; the source code will be made publicly available to facilitate future research and development.
\end{itemize}
The remainder of this paper is organized as follows. Section 2 reviews related work, and Section 3 introduces the detailed system design. Section 4 describes the implementation of the AutoTour mobile application, while Section 5 presents the evaluation results. Section 6 discusses limitations and future work, and Section 7 concludes the paper.
\section{Related Works}
\subsection{Tour Applications}
While a wide range of tour-related applications exist, most exhibit notable limitations in supporting real-time, personalized exploration. For instance, SmartGuide~\cite{smartguide2025}, GetYourGuide~\cite{getyourguide2025}, and Viator~\cite{viator2025} primarily offer static, location-based content and curated experiences but lack dynamic interactivity with the user’s immediate surroundings. Wintor AR Tours~\cite{Wintor2025} leverages augmented reality to deliver on-site narratives, but it requires users to upload personal media (e.g., audio or video), which limits scalability and generalizability. Smartify~\cite{Smartify2025} focuses exclusively on artwork in museums, offering detailed descriptions through a dedicated app, yet it requires custom content design for each specific site. Other tools like Google Lens~\cite{googlelens2025}, AI Tour~\cite{aitour2025}, and Landsnap~\cite{landsnap2025} support image-based landmark recognition but are generally restricted to well-known and iconic locations.

Google Maps’ immersive “Live View” feature introduced in 2023~\cite{googlemapsimmersive2025} overlays annotations such as transit stations, restaurants, and shops onto the user’s camera view, but its coverage is limited to public services and select urban areas where Street View data is available~\cite{googlemapshelp2025}. In contrast to prior work, our system offers a more interactive and engaging way for users to understand their surroundings. Unlike approaches that rely on curated or proprietary content, AutoTour is built on freely available data sources, enabling it to scale to a broader range of cities and regions while maintaining minimal data requirements.
\begin{table}[t!]
\centering
\caption{Comparison of AutoTour and existing tour-related representative applications across key dimensions.}
\label{tab:app:comparison}
\scalebox{0.82}{
\begin{tabular}{c|c|c|c|c|c}
\toprule
\textbf{Applications} & \textbf{Interactivity} & \textbf{Coverage} & \textbf{\begin{tabular}[c]{@{}c@{}}Content\\ Richness\end{tabular}} & \textbf{\begin{tabular}[c]{@{}c@{}}Context\\ Awareness\end{tabular}} & \textbf{\begin{tabular}[c]{@{}c@{}}Open\\ Access\end{tabular}} \\ \midrule
Viator \cite{viator2025} & 
\begin{tabular}[c]{@{}c@{}}\textcolor{red}{\ding{51}}\\ Static content \end{tabular} &
\begin{tabular}[c]{@{}c@{}}\textcolor{ForestGreen}{\ding{51}}\\ Famous features only \end{tabular} &  
\begin{tabular}[c]{@{}c@{}}\textcolor{ForestGreen}{\ding{51}}\\ Visual and audio \end{tabular} &
\begin{tabular}[c]{@{}c@{}}\textcolor{red}{\ding{55}}\\ None \end{tabular}&  
\begin{tabular}[c]{@{}c@{}}\textcolor{red}{\ding{55}}\\ Closed-source \end{tabular} \\ \midrule
Wintor AR Tours~\cite{Wintor2025} &
\begin{tabular}[c]{@{}c@{}}\textcolor{ForestGreen}{\ding{51}\ding{51}\ding{51}}\\ Interactive visual \end{tabular}& 
\begin{tabular}[c]{@{}c@{}}\textcolor{ForestGreen}{\ding{51}}\\ Limited places \end{tabular} &  
\begin{tabular}[c]{@{}c@{}}\textcolor{ForestGreen}{\ding{51}}\\ Visual and audio \end{tabular} &
\begin{tabular}[c]{@{}c@{}}\textcolor{ForestGreen}{\ding{51}}\\ Place aware \end{tabular}&  
\begin{tabular}[c]{@{}c@{}}\textcolor{red}{\ding{55}}\\ Closed-source \end{tabular} \\ \midrule
Google Lens~\cite{googlelens2025} & 
\begin{tabular}[c]{@{}c@{}}\textcolor{ForestGreen}{\ding{51}}\\ Interactive visual \end{tabular}& 
\begin{tabular}[c]{@{}c@{}}\textcolor{ForestGreen}{\ding{51}\ding{51}}\\ Famous features only \end{tabular} &  
\begin{tabular}[c]{@{}c@{}}\textcolor{ForestGreen}{\ding{51}}\\ Text \end{tabular} &
\begin{tabular}[c]{@{}c@{}}\textcolor{ForestGreen}{\ding{51}}\\ Photo aware \end{tabular}&  
\begin{tabular}[c]{@{}c@{}}\textcolor{red}{\ding{55}}\\ Closed-source \end{tabular} \\ \midrule
\begin{tabular}[c]{@{}c@{}}Live View~\cite{googlemapsimmersive2025}\\ (Google Maps)\end{tabular} & 
\begin{tabular}[c]{@{}c@{}}\textcolor{ForestGreen}{\ding{51}\ding{51}}\\ Interactive visual \end{tabular} &
\begin{tabular}[c]{@{}c@{}}\textcolor{ForestGreen}{\ding{51}\ding{51}}\\ Citywide \end{tabular} &
\begin{tabular}[c]{@{}c@{}}\textcolor{ForestGreen}{\ding{51}\ding{51}}\\ Visual and text \end{tabular} &
\begin{tabular}[c]{@{}c@{}}\textcolor{ForestGreen}{\ding{51}\ding{51}\ding{51}}\\ Spatially aware \end{tabular}&  
\begin{tabular}[c]{@{}c@{}}\textcolor{red}{\ding{55}}\\ Closed-source \end{tabular} \\ \midrule
\begin{tabular}[c]{@{}c@{}}\textbf{AutoTour}\\ \textbf{(ours)}\end{tabular}  & 
\begin{tabular}[c]{@{}c@{}}\textcolor{ForestGreen}{\ding{51}\ding{51}\ding{51}}\\ Interactive visual and audio \end{tabular} & 
\begin{tabular}[c]{@{}c@{}}\textcolor{ForestGreen}{\ding{51}\ding{51}\ding{51}}\\ Worldwide \end{tabular} & 
\begin{tabular}[c]{@{}c@{}}\textcolor{ForestGreen}{\ding{51}\ding{51}\ding{51}}\\ Visual, text, and audio \end{tabular} & 
\begin{tabular}[c]{@{}c@{}}\textcolor{ForestGreen}{\ding{51}\ding{51}\ding{51}}\\ Spatially aware \end{tabular} &  
\begin{tabular}[c]{@{}c@{}}\textcolor{ForestGreen}{\ding{51}\ding{51}\ding{51}}\\ Free-source \end{tabular} \\ \bottomrule
\end{tabular}}
\end{table}


\subsection{Contextual Visual Understanding}
\begin{figure}[t!]
    \centering
    \includegraphics[width=0.95\linewidth]{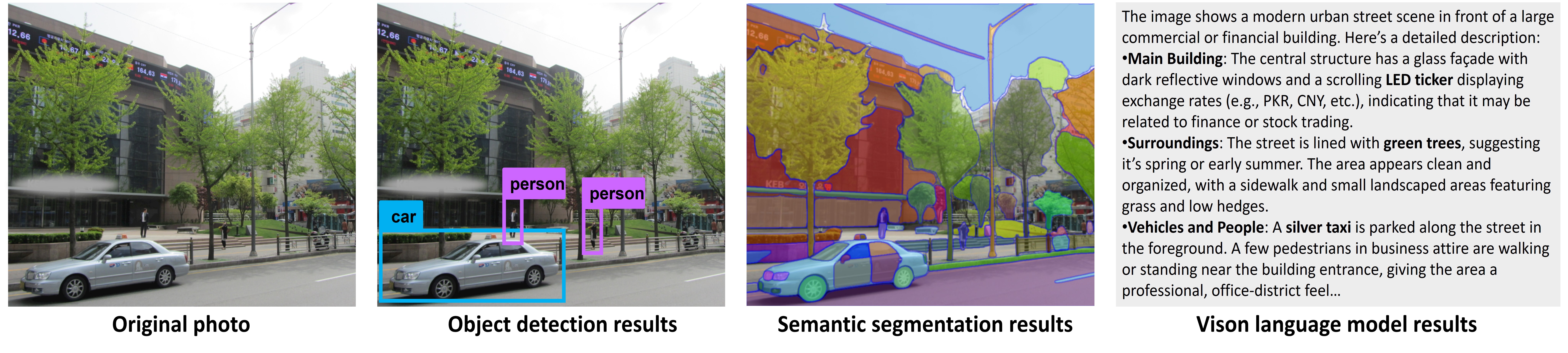}
    \caption{Comparison of existing models with vision capabilities. The semantic segmentation result is from Segment Anything \cite{kirillov2023segany} and the vison language model results are from GPT-4o \cite{openai2025gpt4o}. None of them can achieve the results shown in Fig. \ref{fig:scenario}}
    \label{fig:cvu}
\end{figure}

Recent advances in computer vision have enabled machines to extract increasingly detailed information from visual scenes. Core techniques include object detection (e.g., Faster R-CNN~\cite{ren2015faster}, YOLOv8~\cite{yolov8_2024}), which identifies and localizes objects within an image; and semantic segmentation (e.g., DeepLabv3+~\cite{chen2017rethinking}, SegFormer~\cite{segformer2021}, and Segment Anything \cite{kirillov2023segany}), which classifies each pixel into predefined categories to generate a dense understanding of scene components. However, while these models excel at recognizing generic visual elements, they typically lack the ability to associate those elements with high-level, contextual information such as the historical significance of a building, the name and function of a structure, or region-specific cultural insights.

Recent progress in vision-language models (VLMs) such as GPT-4o~\cite{openai2024gpt4o} and Claude~\cite{anthropic2025claude} has enabled powerful multimodal reasoning capabilities by combining image understanding with natural language processing. These models can interpret visual inputs, answer questions about images, and generate descriptions or captions based on complex visual scenes. However, current VLMs operate primarily through textual interfaces and cannot identify less prominent landmarks or region-specific buildings, especially those absent from major datasets or knowledge corpora. As a result, their outputs often lack the specificity and contextual relevance needed for effective on-site guidance.

The Google Cloud Vision API~\cite{googlemapslandmark2025} offers landmark recognition and global localization capabilities from photos; our empirical tests indicate it performs poorly in identifying less prominent or context-specific features. As such, existing solutions are not well-suited to address the specific task that AutoTour targets: delivering real-time, context-aware, and visually anchored exploration support.

\subsection{LLM-based Sensing}
Large Language Models (LLMs) have achieved remarkable advancements across a wide range of tasks \cite{brown2020language, scao2022bloom, zeng2022glm,openai2023gpt4,touvron2023llama,generalpatternmachines2023} due to their out-of-the-box capabilities trained from large-scale text datasets and ome works \cite{driess2023palm,peng2023kosmos,jiang2022vima,brohan2023rt,ye2023mplug,wang2023visionllm, openai2024gpt4o} extend LLMs into multimodal models. Additionally, several studies introduce innovative LLM applications \cite{liu2023large,xu2024penetrative,yang2024drhouse,arakawa2024prism, chen2024sensor2text, nepal2024mindscape} with sensor data in diverse fields, such as Liu et al.'s work \cite{liu2023large}, which analyzes medical data for health-related tasks. Notably, researchers have proposed the concept of Penetrative AI \cite{xu2024penetrative}, exploring the integration of LLMs with the physical world through IoT sensors. With embedded extensive commonsense knowledge, LLMs/VLMs can perform physical tasks by incorporating IoT sensors. Inspired by the idea of Penetrative AI, we propose a novel application of LLMs/VLMs that senses and interprets users’ surroundings using sensor data collected from smartphones.

\subsection{OpenStreetMap and Overpass API}
\begin{figure} [t!]
    \centering
    \includegraphics[width=1\linewidth]{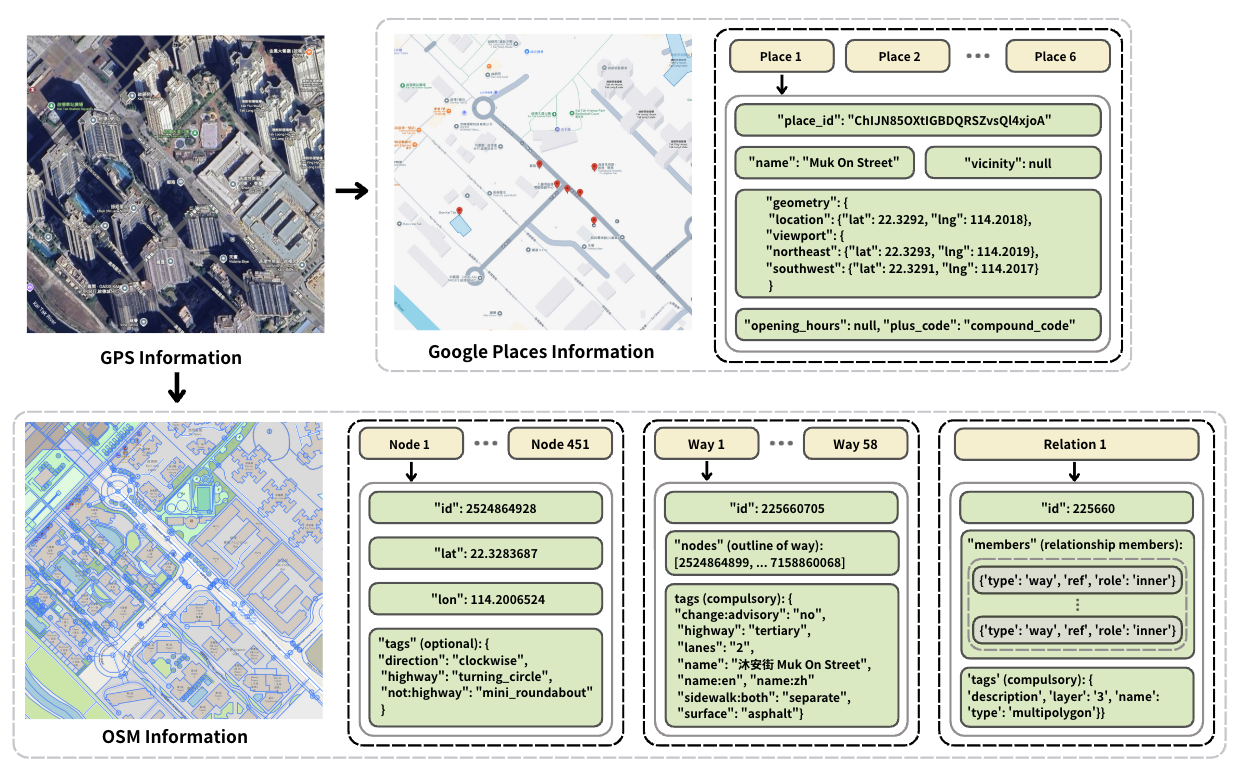}
    \caption{Geospatial database comparison between Google Maps Places API and free-source OSM.}
    \label{fig:osm_google}
\end{figure}

To build AutoTour, we explored the integration of existing geospatial databases that offer comprehensive global coverage. While commercial platforms such as Google Maps Platform~\cite{google_maps_2025} provide APIs like Reverse Geocoding and Places API~\cite{google_places_2025}, the information they return is limited in both quantity and scope. For instance, the Places API can return a small set of nearby locations (typically around six) along with basic attributes such as name, ID, and coordinates, as illustrated in Fig.~\ref{fig:osm_google}. These results are heavily biased toward public-facing venues such as restaurants and retail stores, while many residential buildings and contextually relevant landmarks are often omitted.

In contrast, we find that OpenStreetMap (OSM) is a community-driven geospatial database released under the Open Database Licence (ODbL)~\cite{odbl10} andhas accumulated approximately 10.5 million registered contributors~\cite{osm_stats_2024} and more than 9.9 billion nodes~\cite{taginfo_2025} (as of June 2025), with around four million edits occurring daily~\cite{osm_stats_2024}. More importantly, the dataset is freely available and openly accessible. The Overpass API~\cite{overpass_ql_guide_2025} enables programmatic querying of planet-scale OSM data using a powerful domain-specific language (Overpass QL), making it possible to extract rich and fine-grained contextual information.

OSM data is structured around three core primitives:
\begin{itemize}
    \item \textbf{Nodes} represent individual geographic points (e.g., a tree, bus stop, or mailbox), each defined by latitude and longitude. Nodes may include tags that describe their function or identity, such as \texttt{amenity=bench}, \texttt{natural=tree}, or \texttt{tourism=information}. These tags provide semantic context that can be used to interpret and annotate the element.
    \item \textbf{Ways} are ordered lists of nodes, used to represent linear features (e.g., roads, rivers) or area boundaries (e.g., buildings, parks). Ways may also include descriptive attributes such as \texttt{name}, \texttt{highway}, or \texttt{building}, which specify their type (e.g., \texttt{highway=footway}, \texttt{building=school}) and label (e.g., \texttt{name=Main Street}). These tags help identify the structure and function of the mapped feature.
    \item \textbf{Relations} define logical or geographic relationships among nodes and ways, such as route networks or administrative boundaries. Each member in a relation can have a specific role (e.g., \texttt{outer}, \texttt{inner}, \texttt{stop}, \texttt{platform}), and the relation itself may include tags like \texttt{type=route}, \texttt{type=boundary}, or \texttt{name=Central Line}. This allows OSM to represent complex features such as bus or hiking routes, multi-polygon buildings, or geopolitical borders.
\end{itemize}
As shown in Fig.~\ref{fig:osm_google}, for the same geographic area, OSM provides significantly more comprehensive information than Google Maps, including 451 nodes, 59 ways, and 1 relation. This includes not only location geometry but also descriptive metadata. Therefore, we propose leveraging the OSM database as a reliable and extensible knowledge base for AutoTour.

\section{AutoTour}
\begin{figure}[t!]
    \centering
    \makebox[\textwidth][c]{%
      \includegraphics[width=0.9\textwidth]{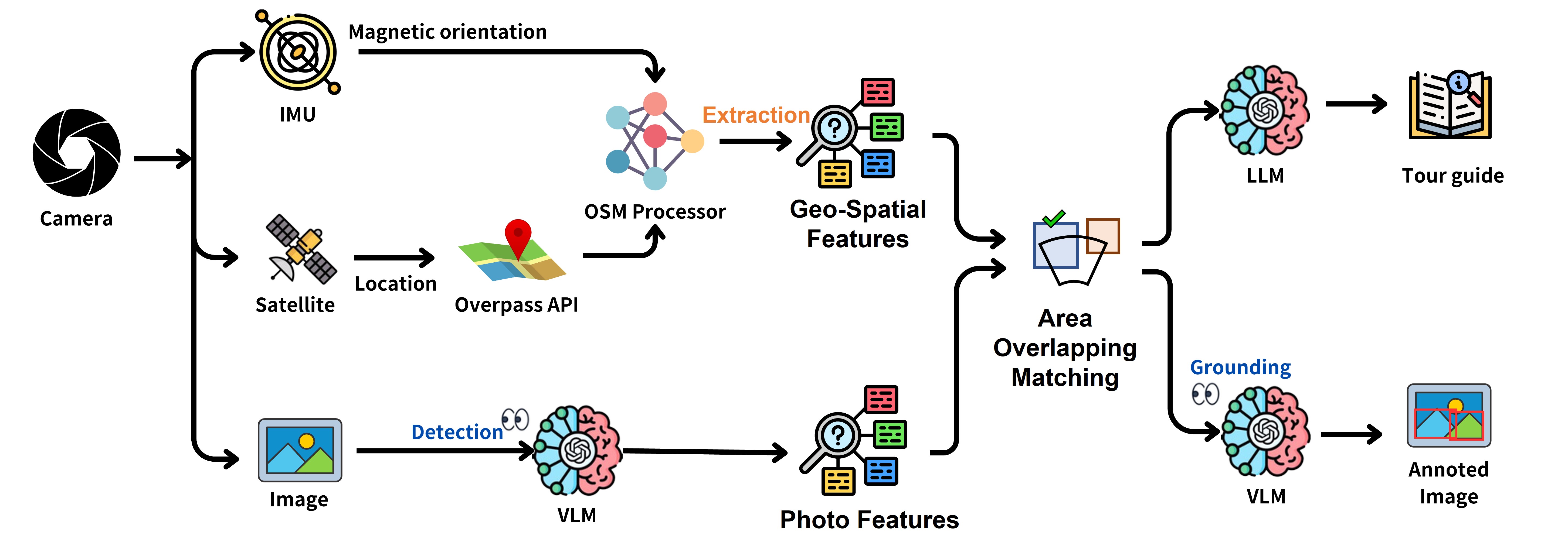}
    }
    \caption{Design workflow of AutoTour.}
    \label{fig:design:main}
\end{figure}
In this section, we present the design of \textit{AutoTour}. An overview of the framework is illustrated in Figure~\ref{fig:design:main}. The system consists of four main components:

\begin{itemize}
    \item \textbf{Geo-Spatial Feature Extraction:}  
    Given the user’s GPS location and the photo shooting direction, this module retrieves surrounding features from the OpenStreetMap (OSM) database and keeps the major features. The extracted elements include points of interest and linear structures (e.g., roads, rivers).

    \item \textbf{Photo Feature Detection:}  
    This module leverages vision-language models (VLMs) to extract features from the user’s photo. It identifies candidate landmarks, buildings, or natural features visible in the captured image.

    \item \textbf{Feature Matching:}  
    The matching module aligns features obtained from OSM with those detected in the photo. The matching leverages both spatial information (e.g., distance, direction) and semantic attributes (e.g., category, name) to match visual elements with corresponding geospatial entities.

    \item \textbf{Presentation:}  
    Finally, the presentation module overlays annotations directly onto the photo, highlighting identified landmarks and features. It also generates context-aware descriptions that can be delivered as text, which can be converted into audio for accessibility.
\end{itemize}

\subsection{Geo-Spatial Feature Extraction}
\begin{figure}[t!]
    \centering
    \makebox[\textwidth][c]{%
      \includegraphics[width=0.85\textwidth]{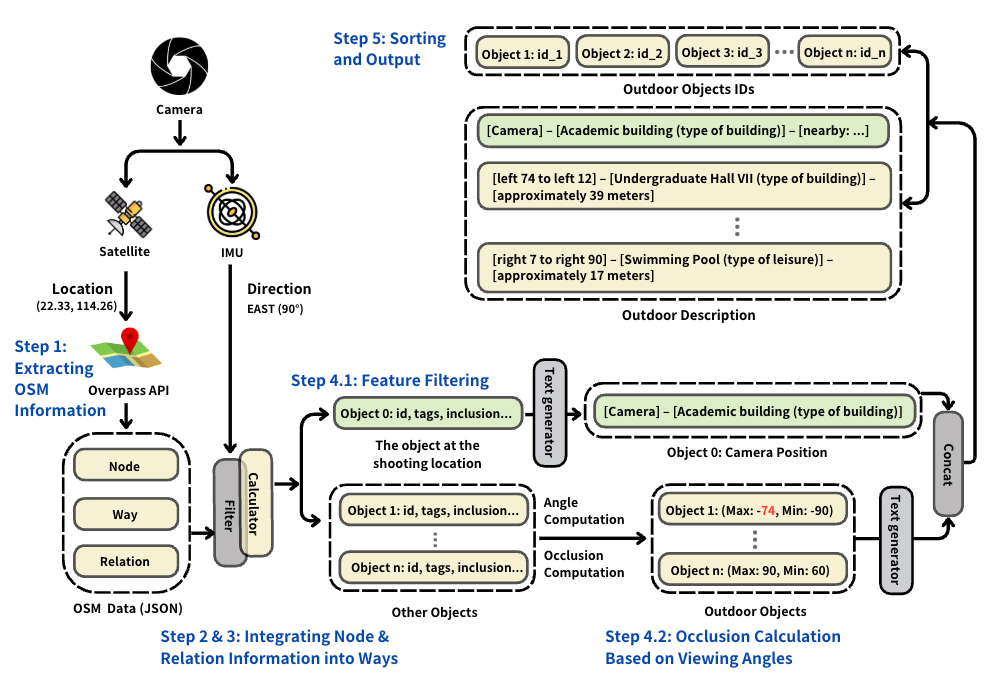}
    }
    \caption{Geo-spatial feature extraction pipeline. Data retrieved from OSM is filtered and structured to retain key attributes for subsequent feature matching with photos.}
    \label{fig:design:gfd}
\end{figure}
We first describe how we preprocess OpenStreetMap (OSM) data retrieved via the Overpass API to prepare features for subsequent matching. Although raw OSM data is comprehensive, including detailed attributes of nodes, ways, and relations, it also contains large amounts of redundant information (e.g., internal IDs) that are not directly useful for feature matching. Therefore, we design a three-step feature extraction pipeline, illustrated in Fig.~\ref{fig:design:gfd}, to retain only relevant and structured attributes.

\textbf{Granularity and format unification}. We begin by retrieving all nearby nodes, ways, and relations around the photo’s GPS location via the Overpass API. Since these primitives differ in granularity, we unify them as follows:
\begin{itemize}
    \item \textbf{Nodes to Ways}: Nodes typically represent small features (e.g., a shop, bus stop). To avoid fragmented representation, we embed nodes into larger geographic structures (ways) based on spatial proximity and logical relations. This ensures that fine-grained entities are contextualized within higher-level features such as streets, parks, or buildings.
    \item \textbf{Relations to Ways}: Relations often describe large or composite structures (e.g., a block formed by several ways, or a bus route linking multiple stops). To simplify their representation, we associate relations with corresponding ways. This consolidation yields a more holistic, coherent view of the environment.
\end{itemize}
Through this step, both nodes and relations are transformed into ways, producing a consistent granularity and unified representation.

\textbf{Feature filtering}. Since photos capture only the visible subset of the environment, we filter features according to the user’s perspective. We first calculate the field of view based on the photo’s GPS position and shooting direction, retaining only features within the angular range of visibility. We then apply occlusion filtering, estimating whether closer objects obstruct those farther away. This step simulates a realistic human visual perspective, ensuring that the retained features align with what the user is likely to see in the photo.

\textbf{Feature reorganization}. In the final step, we reorganize the filtered features into a concise and structured format that facilitates comparison with photo-detected features. Each feature entry consists of several components:
\begin{enumerate}
    \item \textbf{Geo-coordinates} -- the geographic coordinate data (latitude and longitude) of the points that constitute the feature on the map, such as the vertices of a building footprint, the centerline of a road segment, or the boundaries of a park area.
    \item \textbf{Name} -- the designated identifier or label of the feature, such as the official name of a building, road, or landmark.
    \item \textbf{Inclusive Information} -- describes elements or attributes contained within the feature area, including internal structures, facilities, or related metadata.
    \item \textbf{Nearby Information} -- lists adjacent or surrounding features that provide spatial or contextual relevance for navigation and recommendation.
    \item \textbf{Categorizer} -- specifies the feature type, such as \textit{building}, \textit{road}, \textit{park}, or other thematic classes, enabling efficient organization and filtering.
\end{enumerate}

All features are then sorted by angle range (left to right) to mirror the user’s visual perspective. This final representation is concise and semantically rich for the future feature matching process.

\subsection{Photo Feature Detection}

\begin{figure}[htbp]
    \centering
    \makebox[\textwidth][c]{%
      \includegraphics[width=0.8\textwidth]{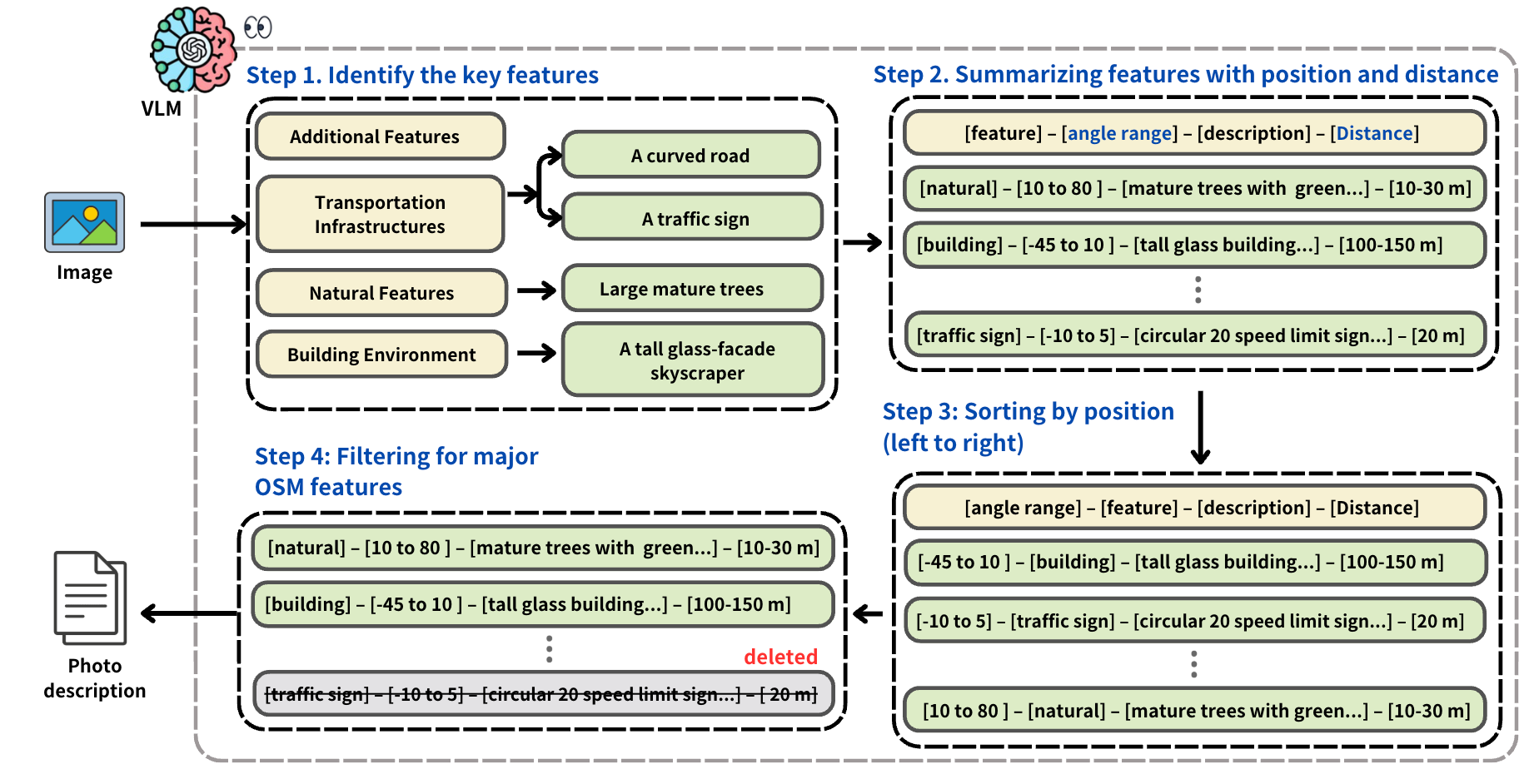}
    }
    \caption{Pipeline of photo feature detection using vision-language models for extracting structured visual features.}
    \label{fig:design:pfd}
\end{figure}

Next, we extract features from photos. We first surveyed existing object detection models such as Faster R-CNN~\cite{ren2015faster} and YOLO~\cite{yolov8_2024}, but found that they rarely perform well on landmarks or buildings. Most of these models focus on detecting people or vehicles, and there are no large-scale building detection datasets available for training or fine-tuning related models. 

To address this, we turn to recent vision-language models (VLMs), such as GPT-4o, which demonstrate strong image understanding and can capture most features while also generating detailed descriptions of them. However, directly prompting VLMs for feature detection does not provide optimal performance. Therefore, we design a procedure to enhance their outputs by constraining formats, incorporating spatial priors, and filtering results, which makes the extracted features more reliable for the following matching step.

We first tested a naive prompt—``Describe this image''—with GPT-4o and observed that although the model successfully recognized key visual elements, it often generated trivial objects such as flowers or streetlights that fall outside the main objectives of our system. Moreover, it sometimes produced hallucinated items—two of which did not actually exist in the image (i.e., false positives). In addition, the generated descriptions were written in free-form text, lacking structure and consistency, which makes them unsuitable for systematic feature matching. 

To address the challenge and improve the precision and consistency of feature extraction, we designed a structured, multi-step prompting procedure for the vision-language model as shown in Fig. \ref{fig:design:pfd}:

\begin{itemize}
    \item \textbf{Step 1: Category Guidance.}  
    Identify key features within four major categories---\textit{transportation infrastructure, natural features, built environment, and additional landmarks}. This focuses the model on dominant OSM classes while suppressing long-tail objects (e.g., lampposts, benches).

    \item \textbf{Step 2: Structured Description.}  
    Summarize each detected feature in the format:  
    \textit{[feature name] -- [angle span] -- [description] -- [distance]}.  
    This ensures the output includes spatial and semantic details necessary for subsequent feature matching.

     \item \textbf{Step 3: Left-to-Right Ordering.}  Sort all detected features according to their left-to-right position in the image. This provides the downstream map-matching module with a deterministic sequence that aligns with the photographic azimuth.

    \item \textbf{Step 4: Category Alignment.}  
    Retain only features aligned with core OSM keys:  
    \{\texttt{building}, \texttt{road}, \texttt{park}, \texttt{natural}, \texttt{waterway}\}.  
    This step filters out irrelevant or trivial objects (e.g., flowers, laptops) that are not the major purpose of our system and improves the model’s correspondence with OSM’s data taxonomy.
\end{itemize}

We test our prompt with a representative street-level photograph as shown in Fig. \ref{fig:vlm_prompt} along with several minor artefacts such as traffic lights, bollards, and signage. We compared two prompting strategies:

\begin{figure}[t!]
    \centering
    \makebox[\textwidth][c]{%
      \includegraphics[width=0.4\textwidth]{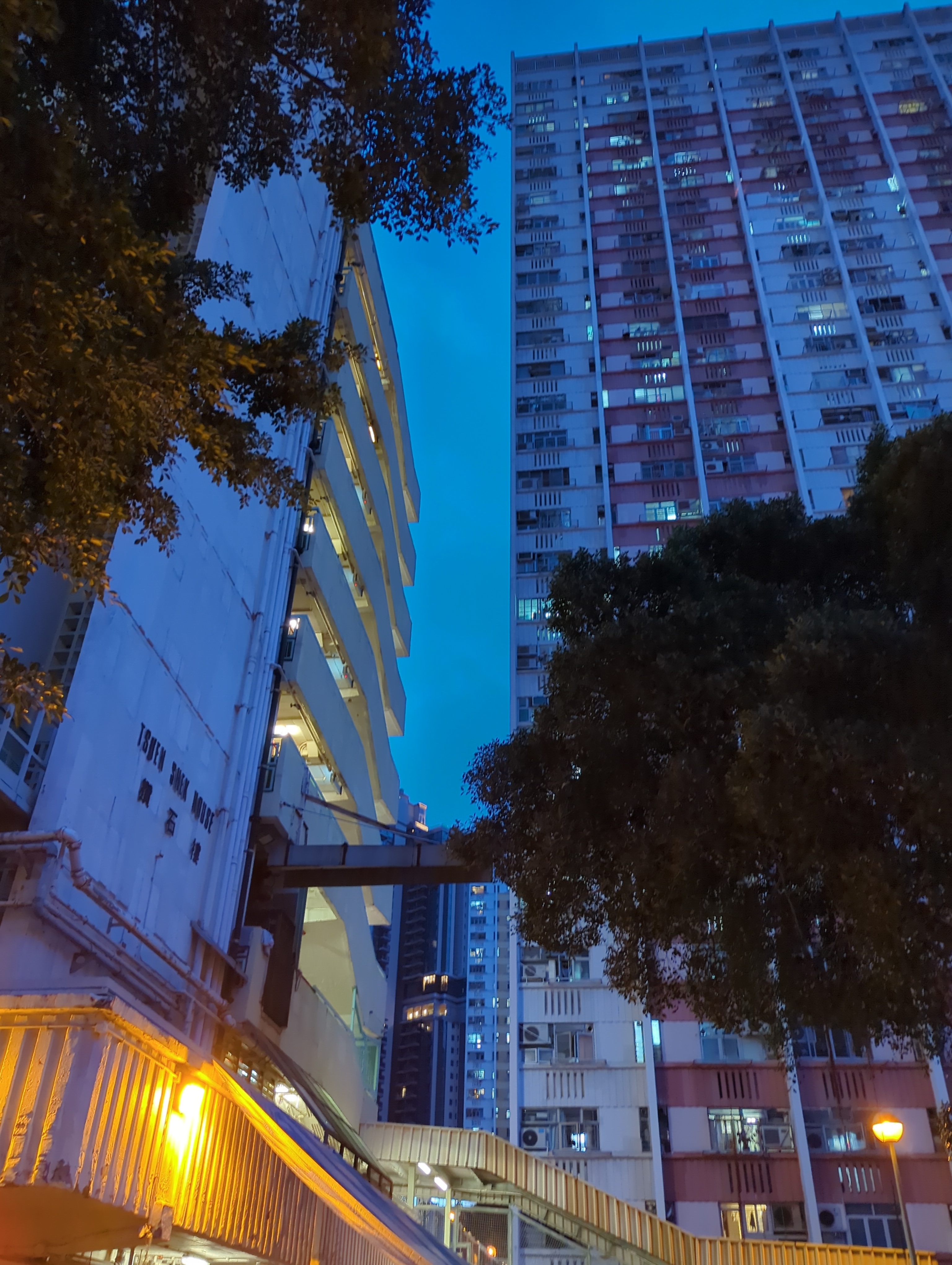}
    }
    \caption{Urban scene image used for testing photo feature detection prompt.}
    \label{fig:vlm_prompt}
\end{figure}

\begin{enumerate}
    \item {\textbf{Baseline prompt:} \emph{``Describe this image.''}}
    \begin{itemize}
        \item {\textbf{Result:} 
            \begin{quote}
            \small
            The image depicts an urban scene during twilight or early evening. Two tall residential \textbf{buildings} dominate the scene, one on the left and another on the right. The building on the left is white and has a sign with text on it, while the building on the right is taller and has a mix of white and red colors. Both buildings have numerous windows, many of which are illuminated, indicating that people are inside.  
        
            In the foreground, there is a\textbf{ pedestrian bridge or walkway} with a yellow-orange light illuminating it. \textbf{Trees} with dark foliage are visible, partially obscuring parts of the buildings. The \textbf{sky} is a deep blue, suggesting that it is either dusk or early evening. The overall atmosphere is a mix of urban density and evening calm.
            \end{quote} }
        \item{\textbf{Analysis:} The model returned \textbf{5} candidate items, of which \textbf{3} were true-positive OSM objects and \textbf{2} were false positives. }
    \end{itemize}
    \item {\textbf{Our prompt:} described in \S 3.2.1.}
    \begin{itemize}
        \item {\textbf{Result:} 
            \begin{quote}
                \small
                [left 70° to left 30°] — [Multi-storey building (left)] — [White building with balconies] — [$\sim$20 m] \newline
                [left 10° to right 10°] — [Elevated walkway] — [Pedestrian bridge with a roof] — [$\sim$5–20 m]      \newline   
                [right 30° to right 70°] — [Multi-storey building (right)] — [Tall building with red/white façade] — [$\sim$30 m]
            \end{quote} }
        \item{\textbf{Analysis:} Exactly \textbf{3} items were returned, all of which matched ground-truth OSM entities; no false positives were produced. Because the output already satisfied positional and semantic constraints, no post-processing was required, and overall inference time fell by approximately 20\%.}
    \end{itemize}

\end{enumerate}

\subsection{Feature Matching}
\label{sec:feature:matching}

\begin{figure}[ht]
\centering
\makebox[\textwidth][c]{%
\includegraphics[width=1.0\textwidth]{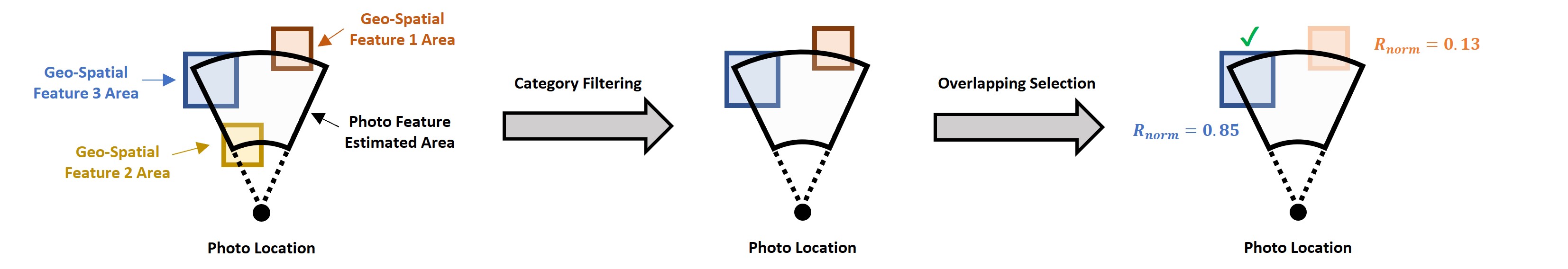}
}
\caption{Geometric-based feature matching algorithm. The semi-annular sector indicates the estimated area of a photo feature defined by the angle and distance ranges from the VLM. Geo-spatial feature 2 is selected as the match because it has the largest overlap compared to its own area.}
\label{fig:matching}
\end{figure}
After obtaining the features and their detailed attributes from both OpenStreetMap (OSM) and the detected photo content, we proceed to map them together through a geometry-driven fusion approach that evaluates the spatial overlap between photo-detected and geo-spatial features as shown in Fig. \ref{fig:matching}.

During photo analysis, the VLM estimates, for each detected feature, its approximate \emph{angle range} and \emph{distance range} relative to the photo location. These parameters define a \emph{semi-annular sector}—a half-ring–shaped region centered at the photo’s capture location. This sector represents the estimated spatial extent where the corresponding feature is likely situated on the ground plane.

From the map side, we retrieve the \textbf{geo-coordinates} of nearby geo-spatial features, such as building footprints, roads, or park boundaries. To establish correspondence between photo-derived and map-based features, we proceed as follows:

\begin{enumerate}
    \item \textbf{Category Filtering:} 
    First, we filter geo-spatial features by category (e.g., buildings, parks) using the category of photo feature:
    \[
        F_{\text{filtered}} = \{ f_i \mid \text{Category}(f_i) \in \text{PhotoFeatureCategories} \}.
    \]
    \item \textbf{Overlap Computation:} 
    For each photo-detected feature and the corresponding semi-annular sector, we compute the geometric intersection with filtered geo-spatial feature polygons:
    \[
        A_{\text{overlap}} = \text{Area}(S_{\text{photo}} \cap P_{\text{map}}),
    \]
    where $S_{\text{photo}}$ is the semi-annular sector and $P_{\text{map}}$ is the polygon of the map feature.
    \item \textbf{Normalized Overlap Ratio:} 
    We then evaluate the relative overlap using the normalized ratio:
    \[
        R_{\text{norm}} = \frac{A_{\text{overlap}}}{A_{\text{map}}},
    \]
    where $A_{\text{map}}$ is the total area of the map feature polygon.
    \item \textbf{Feature Matching:} 
    Finally, the Geo-Spatial feature with the largest normalized overlap ratio is selected as the matched geographic entity:
    \[
        f_{\text{matched}} = \arg\max_{f_i \in F_{\text{filtered}}} R_{\text{norm}}(f_i).
    \]
\end{enumerate}
If a photo-detected feature fails to match any map feature—such as in cases where the map lacks the corresponding object—its information is preserved independently. 
These unmatched features remain valuable, as they still represent important visual entities in the photo. 
For such cases, we directly retain the descriptive output provided by the VLM, ensuring that all meaningful visual observations are included in the final output, even when a geographic correspondence cannot be established.

\subsection{Feature Presentation}

This section outlines the procedure for presenting detected features to users in two complementary ways: \textit{visual} and \textit{textual}.

\begin{figure}[htbp]
\centering
\makebox[\textwidth][c]{%
\includegraphics[width=0.8\textwidth]{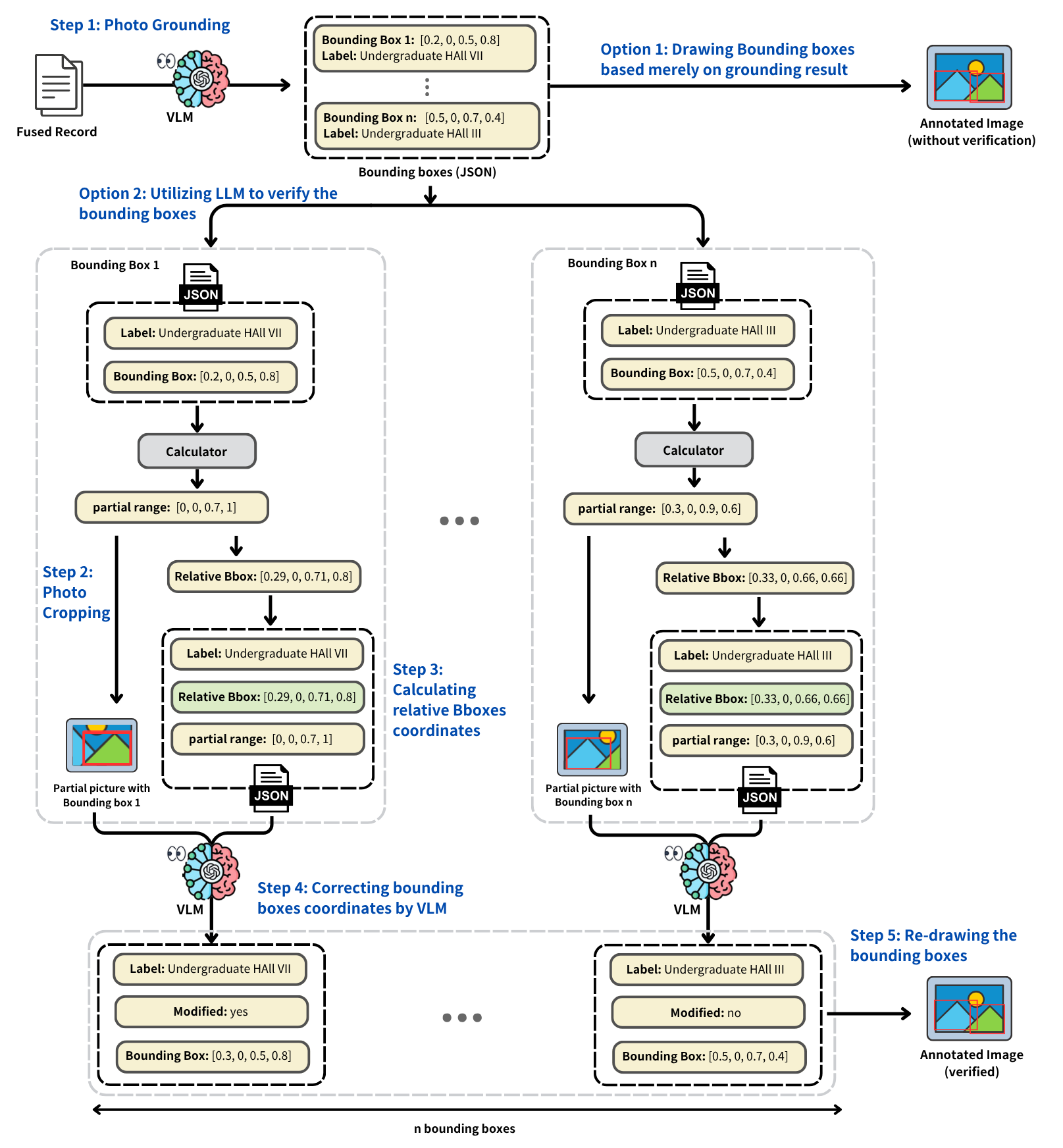}
}
\caption{Feature Visualization and Bounding boxes fixing}
\label{fig:design3}
\end{figure}

\subsubsection{Visual Display}\label{sec:design:visual}
To facilitate direct and intuitive interaction, we annotate the detected features back onto the original photos so that users can clearly identify them in context.  To achieve this, we leverage the \texttt{qwen-vl-max} model~\cite{Qwen2.5-VL}, which supports grounding---that is, it can take an image and a textual description as input and return the bounding box (\texttt{bbox}) of the object corresponding to the text. An example output is shown below:

\begin{verbatim}
{
  "label": "High-rise buildings (left side)",
  "bounding_box": [0.0, 0.0, 0.62, 1.0]
}
\end{verbatim}

To reduce system overhead, each bounding box is then used to crop the corresponding region from the original image. Only these cropped regions are processed further for analysis and annotation, which enables more focused evaluation of each feature. For storage and downstream alignment, we record the \textit{relative bounding box coordinates} and \textit{crop ranges} in JSON format, as illustrated below:

\begin{verbatim}
{
  "label": "High-rise buildings (left side)",
  "bounding_box": [0.0, 0.0, 0.62, 1.0],
  "crop_range": [0.0, 0.0, 0.52, 1.0]
}
\end{verbatim}

However, we also observed that the grounding model may produce inaccurate results since the model was not explicitly trained for architectural annotation. To improve grounding accuracy, we introduce an additional post-processing step to refine the bounding boxes using a vision-language model. Specifically, we present the original photo together with the draft bounding boxes generated by \texttt{qwen-vl-max} to a secondary model, which judges whether each bounding box adequately covers the corresponding feature. If not, the model outputs updated bounding box coordinates. To further reduce processing time, each bounding box correction is performed in parallel using the vision-language model, significantly improving efficiency without compromising accuracy. The revised data, including both the original and modified bounding boxes, is stored in a structured JSON format, as shown below:\begin{verbatim}
{
  "label": "High-rise buildings (right side)",
  "modified": "no",
  "bounding_box": [0.27, 0.0, 1.0, 1.0]
}
\end{verbatim}

Finally, the corrected bounding box coordinates are applied to the original image. Using the modified coordinates, we redraw the boxes to produce an updated visualization where the feature regions are accurately delineated. This step enhances spatial precision and ensures that visual annotations align closely with the true scene content.

\subsubsection{Textual Display}
In addition to the visual presentation, we also generate textual descriptions for each matched feature. To achieve this, we design a prompt that includes all detected and matched features, and instruct the LLM to compose a narrative-style explanation that functions as a virtual tour guide for the user. The generated text summarizes key details such as feature names, types, historical or cultural relevance, and spatial relationships between landmarks.

To enhance accessibility and engagement, the textual descriptions can be further converted into speech using text-to-speech (TTS) techniques. This allows users to receive real-time auditory guidance, facilitating hands-free exploration and a more immersive interaction experience.

\section{Autotour APP}
\begin{figure}[t!]
\centering
\makebox[\textwidth][c]{%
\includegraphics[width=0.9\textwidth]{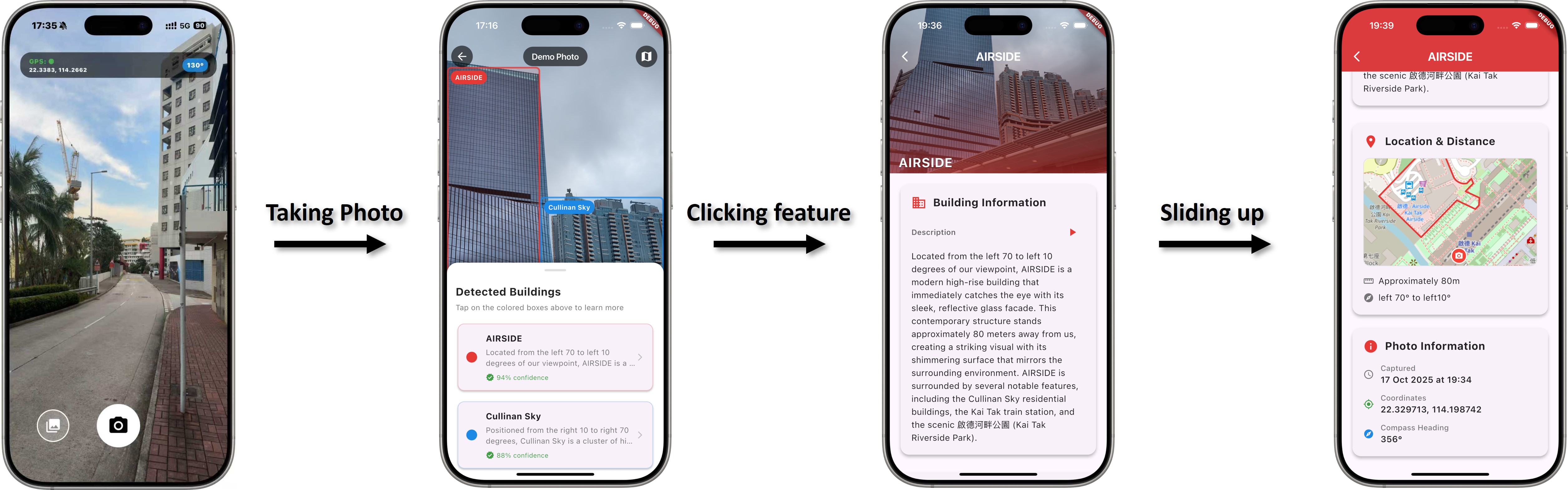}
}
\caption{Key user interface of the AutoTour application.}
\label{fig:app}
\end{figure}
We also develop a working prototype of \textit{AutoTour}. This section summarizes the core user-facing screens in the mobile application and the key interactions supported on each.

\subsection{Core User Interfaces}
\begin{itemize}
    \item \textbf{Main Camera:} The primary entry point for photo capture with contextual sensing. The interface features a full-screen live camera preview, a compact GPS and heading indicator, a central shutter button, and quick access to the gallery. Users can tap once to capture a photo or open the gallery to review previous captures.

    \item \textbf{Photo Analysis:} After a photo is taken, the app presents a smooth transition animation before proceeding to this interface. The screen displays detection and annotation results directly over the captured photo, with color-coded bounding boxes and a draggable bottom sheet listing identified features with concise descriptions. A play icon enables text-to-speech narration. Users can tap a bounding box or list item to view detailed information about the corresponding building or landmark.

    \item \textbf{Building Details:} This screen provides detailed information about a selected landmark or building. A hero header highlights the selected region within the original image, while a text-to-speech function narrates the building’s description. Users can slide down to access the \textit{Location \& Distance} card.
\end{itemize}

\subsection{Implementation}

The prototype of \textit{AutoTour} is implemented as a cross-platform mobile application designed for both Android and iOS devices.

\textbf{Mobile Client.}  
The mobile client is developed using the Flutter framework (Material 3 design system), which enables a unified codebase for Android and iOS deployment. We adopt \textit{Riverpod} for application-level state management. Riverpod offers predictable and reactive state handling, enabling efficient coordination between asynchronous components such as camera input, GPS sensor updates, and server-side analysis results. To minimize latency, network-intensive tasks (e.g., photo uploads, OSM queries, model inference) are executed asynchronously with user-visible progress indicators. 

\textbf{Server.}
We deploy a stateless, cloud-ready service that bridges the mobile client with the fusion pipeline via a lightweight REST interface. The server (FastAPI + Uvicorn) accepts a photo with GPS/heading metadata, enqueues an asynchronous background job, and exposes minimal health, status, and result retrieval primitives—omitting low-level details here to avoid repetition. It emphasizes geometry-first evidence while remaining deployment-friendly: concurrency-safe processing, optional realtime progress upserts (e.g., Supabase), and pluggable storage for jobs/artifacts in production. 

\textbf{Model configuration}. Given the wide variety of available VLMs, we leverage existing ones to establish a default configuration for \textit{AutoTour}. Specifically, we use \textit{Claude}~\cite{anthropic_claude} for photo feature detection and \textit{Qwen-VL-Max}~\cite{Qwen2.5-VL} for visual grounding and annotation display. We also experiment with several alternative VLMs, and a detailed comparison of their performance can be found in Section~\ref{sec:eva}.
\section{Evaluation}\label{sec:eva}
\subsection{Dataset Collection}
We collected a custom dataset using an Android application developed for this study. The app allows users to capture photos along with their corresponding GPS coordinates and shooting directions, derived from the device’s built-in IMU sensors. We recruited 35 volunteers and conducted data collection across five cities—Hong Kong, Beijing, Shanghai, Shenzhen, and Los Angeles. In total, 134 photos were captured, covering a wide range of urban and suburban environments under both daytime and nighttime conditions.  

\begin{figure}[t!]
\hspace{0.5cm}
  \centering
  \begin{minipage}[b]{0.3\textwidth}
    \centering
    \includegraphics[width=\linewidth]{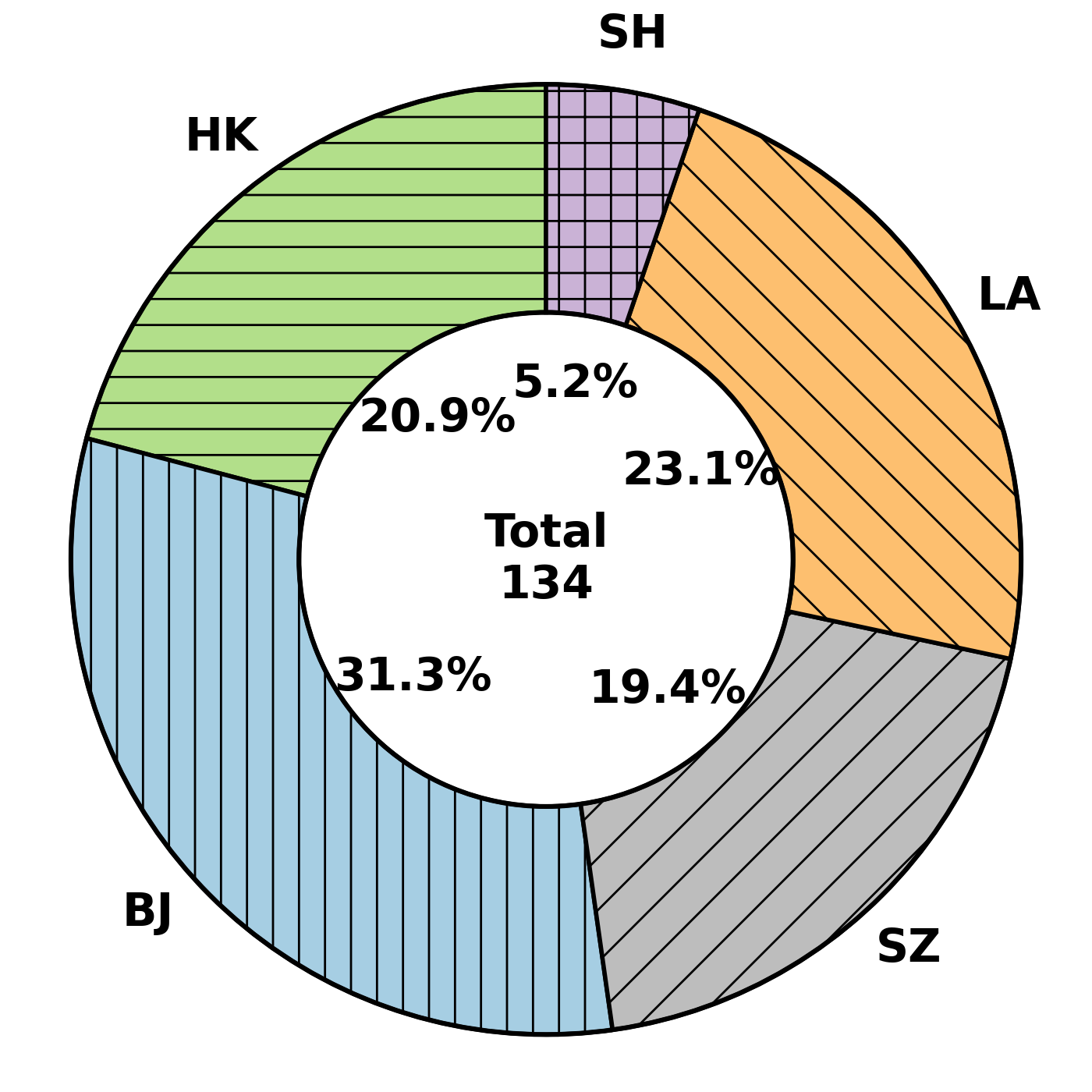}
    \caption{Distribution of evaluation samples across five cities, including Beijing (VJ), Shanghai (SH), Hong Kong (HK), Los Angeles (LA).}
    \label{fig:city_distribution}
  \end{minipage}
  \hfill
  \begin{minipage}[b]{0.5\textwidth}
    \centering
    \includegraphics[width=\textwidth]{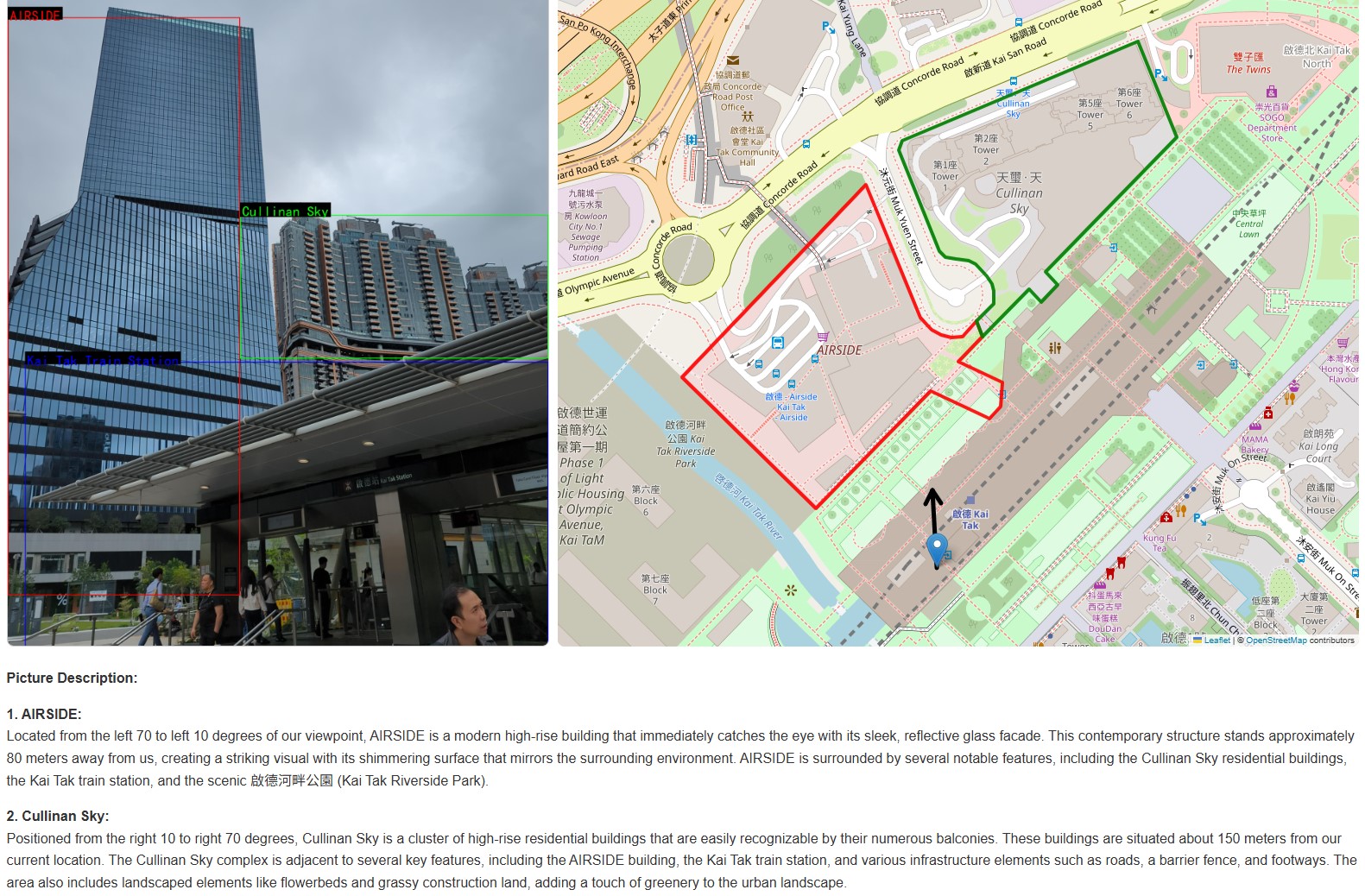}
    \caption{Example of AutoTour results, including detected photo features, matched OSM elements, and generated feature descriptions for user evaluation.}
    \label{fig:eva:web}
  \end{minipage}
  \hspace{0.5cm}
\end{figure}

\subsection{Procedure} To evaluate the performance of AutoTour, we developed a web-based evaluation platform where users can assess the accuracy and completeness of the system’s landmark detection and matching results. The goal of the evaluation is to determine how effectively the system aligns detected features in photos with corresponding elements from OpenStreetMap (OSM).

For every example, the interface displays two images and one textual description as shown in Fig. \ref{fig:eva:web}:
\begin{itemize}
    \item \textbf{Left (Real Photo):} The original photo captured by the user, with bounding boxes drawn around the detected objects.
    \item \textbf{Right (GIS Map):} The corresponding map region centered on the shooting location, with an arrow indicating the camera’s viewing direction.
    \item \textbf{Description:} A concise text summary of the detected objects in the photo that can also be observed on the map.
\end{itemize}
Participants are instructed to review each example and answer four evaluation questions:
\begin{enumerate}
    \item \textbf{Completeness of identification:} Whether all major buildings visible in the photo are correctly detected, and if any important structures are missing or irrelevant ones are included.
    \item \textbf{Correctness of matching:} Whether each detected building corresponds to the correct element on the OSM map.
    \item \textbf{Annotation accuracy:} Whether the bounding boxes precisely enclose the intended features, accurately reflecting their positions and boundaries.
    \item \textbf{Text description accuracy:} Whether the textual description of each building is clear and consistent with its visual and spatial characteristics.
\end{enumerate}

Our project is reviewed and approved by our Institutional Review Board (IRB) under a Human Research Ethics Protocol. Informed consent was obtained from all participants involved before the data collection and evaluation process. Appropriate safeguards were implemented to anonymize and securely store the collected data.

\subsection{Metrics}

To comprehensively evaluate our system, we categorize the employed metrics into two groups: (1) \textbf{User Study Metrics}, used to assess the overall output quality and user-centered results; and (2) \textbf{System Metrics}, focusing on the core technical components of photo feature detection and feature matching. 

\subsubsection{User Study Metrics}\label{sec:metric:user}

These metrics reflect the overall output quality and the effectiveness of user-facing results.

\begin{itemize}
    \item \textbf{Completeness (C):} The ratio of correctly detected major buildings to the total number of major buildings visible in the photo.
    \[
    C = \frac{N_{\text{correctly identified}}}{N_{\text{ground truth}}}
    \]
    \item \textbf{Matching Accuracy (M):} The proportion of correctly matched photo features to their corresponding OSM elements.
    \[
    M = \frac{N_{\text{correct matches}}}{N_{\text{total matches}}}
    \]
    \item \textbf{Bounding Box Precision (B):} The ratio of correctly defined bounding boxes by the user to the total number of bounding boxes.
    \[
    B = \frac{N_{\text{correct bounding box}}}{N_{\text{total boxes}}}
    \]
    \item \textbf{Description Quality (D):} A score evaluating the correctness and informativeness of the generated descriptions.
\end{itemize}
All metrics are rated on a 1–4 scale by users, where 4 indicates well-covered or highly accurate results, 3 indicates mostly covered or accurate results, 2 represents partially covered or accurate results, and 1 denotes poor or inaccurate coverage.
We also report an overall performance score computed as a weighted average:
\[
S = \alpha C + \beta M + \gamma B + \delta D
\]
where $\alpha$, $\beta$, $\gamma$, and $\delta$ are weighting coefficients determined empirically based on importance (default: 0.3, 0.3, 0.2, 0.2).

\subsubsection{System Metrics}

These metrics are specifically designed to evaluate the core modules discussed in Section~5.5.1 \textit{Photo Feature Detection} and Section~5.5.2 \textit{Feature Matching}. They quantify the detection and matching performance of the system in terms of \textbf{Recall and Precision}.

\begin{itemize}
    \item \textbf{Identification Recall and Precision (IR, IP):} These metrics measure the ability of the system to correctly identify major buildings within the input photo. Recall reflects the proportion of correctly detected major buildings relative to all visible ones, while Precision indicates the proportion of correctly detected buildings among all identified candidates.
    \[
    IR = \frac{N_{\text{correctly identified}}}{N_{\text{ground truth}}}, \quad
    IP = \frac{N_{\text{correctly identified}}}{N_{\text{totally identified}}}
    \]
    \item \textbf{Matching Recall and Precision (MR, MP):} These metrics assess the correctness of feature matching to corresponding OSM elements. Recall quantifies the proportion of ground truth features that are successfully matched, and Precision measures the proportion of matches that are correct.
    \[
    MR = \frac{N_{\text{correct matches}}}{N_{\text{ground truth}}}, \quad
    MP = \frac{N_{\text{correct matches}}}{N_{\text{totally matches}}}
    \]
\end{itemize}

\subsection{User Study Evaluation}


\begin{table}[t!]
\centering
\caption{Average scores (1–4) of four evaluation metrics per city, and a weighted overall score 
($S = 0.3C + 0.3M + 0.2B + 0.2D$).}
\label{tab:eva:city}
\scalebox{0.95}{
\begin{tabular}{c|c|c|c|c|c|c}
\toprule
\textbf{Cities} & \textbf{Beijing} & \textbf{Hong Kong} & \textbf{Shanghai} & \textbf{Shenzhen} & \textbf{Los Angeles} & \textbf{Average} \\ \midrule
\textbf{Completeness Mean (↑)}  & 3.908 & 3.558 & 3.857 & 3.766 & 3.560 & 3.730 \\
\textbf{Completeness SD (↓)}    & 0.289 & 0.497 & 0.350 & 0.457 & 0.496 & -- \\ \midrule
\textbf{Matching Accuracy Mean (↑)}  & 3.892 & 3.538 & 3.667 & 3.667 & 3.616 & 3.676 \\
\textbf{Matching Accuracy SD (↓)}    & 0.296 & 0.498 & 0.471 & 0.471 & 0.487 & -- \\ \midrule
\textbf{Bounding Box Precision Mean (↑)}  & 3.538 & 3.308 & 2.667 & 3.450 & 2.998 & 3.192 \\
\textbf{Bounding Box Precision SD (↓)}    & 0.498 & 0.689 & 0.944 & 0.666 & 0.867 & -- \\ \midrule
\textbf{Description Quality Mean (↑)}     & 3.769 & 3.635 & 3.667 & 3.639 & 3.546 & 3.651 \\
\textbf{Description Quality SD (↓)}       & 0.421 & 0.481 & 0.471 & 0.480 & 0.507 & -- \\ \midrule
\textbf{Weighted Score (↑)} & 3.798 & 3.516 & 3.491 & 3.641 & 3.447 & 3.579 \\ \bottomrule
\end{tabular}}
\end{table}

\subsubsection{Overall performance}.  We present the overall performance scores of AutoTour across five cities—\textbf{Beijing (BJ)}, \textbf{Shanghai (SH)}, \textbf{Shenzhen (SZ)}, \textbf{Hong Kong (HK)}, and \textbf{Los Angeles (LA)}—as shown in Table~\ref{tab:eva:city}. These cities exhibit distinct characteristics: for example, Shenzhen and Hong Kong feature high building densities, while Los Angeles and Beijing are relatively sparse. Shenzhen’s landmarks also tend to follow a more modern architectural style. Each city subset includes a balanced number of photo–OSM pairs representing diverse urban contexts, such as dense historic centers, waterfront areas, and open road networks.
For each city, we computed an overall performance score using a weighted combination of completeness, matching accuracy, bounding box precision, and description quality, as defined in Section~\ref{sec:metric:user}. The results show that AutoTour consistently achieves high scores (above 3.0) across most metrics with a total average score of 3.579, demonstrating strong generalizability across different urban environments.

\begin{figure}[t!]
\centering
\makebox[\textwidth][c]{%
\includegraphics[width=1.0\textwidth]{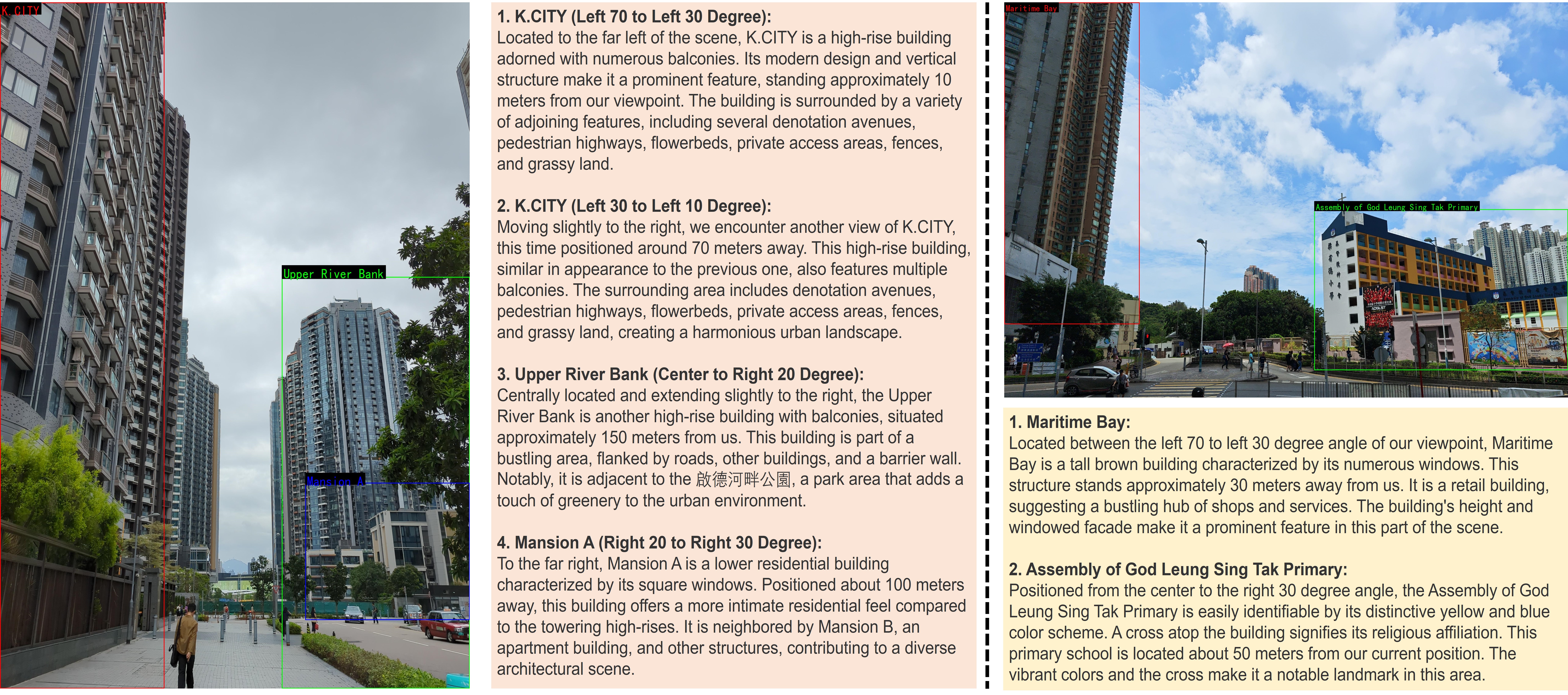}
}
\caption{Two representative outputs of AutoTour in Shenzhen and Hong Kong. Each image presents a photo with automatically detected and annotated features, accompanied by text descriptions generated by the system.}
\label{fig:show}
\end{figure}
Fig. \ref{fig:show} demonstrates two examples of AutoTour’s generated results, including annotated photos with detected features, their names, and overall scene descriptions. In both cases, AutoTour successfully identifies most major landmarks or buildings and provides their correct names. The accompanying text further offers detailed descriptions of the detected features. More results show that it can also detect features like lakes and courts.

\subsubsection{Impact of feature distance}

\begin{figure}[H]
  \centering
  \begin{minipage}[b]{0.45\linewidth}
    \centering
    \includegraphics[width=\linewidth]{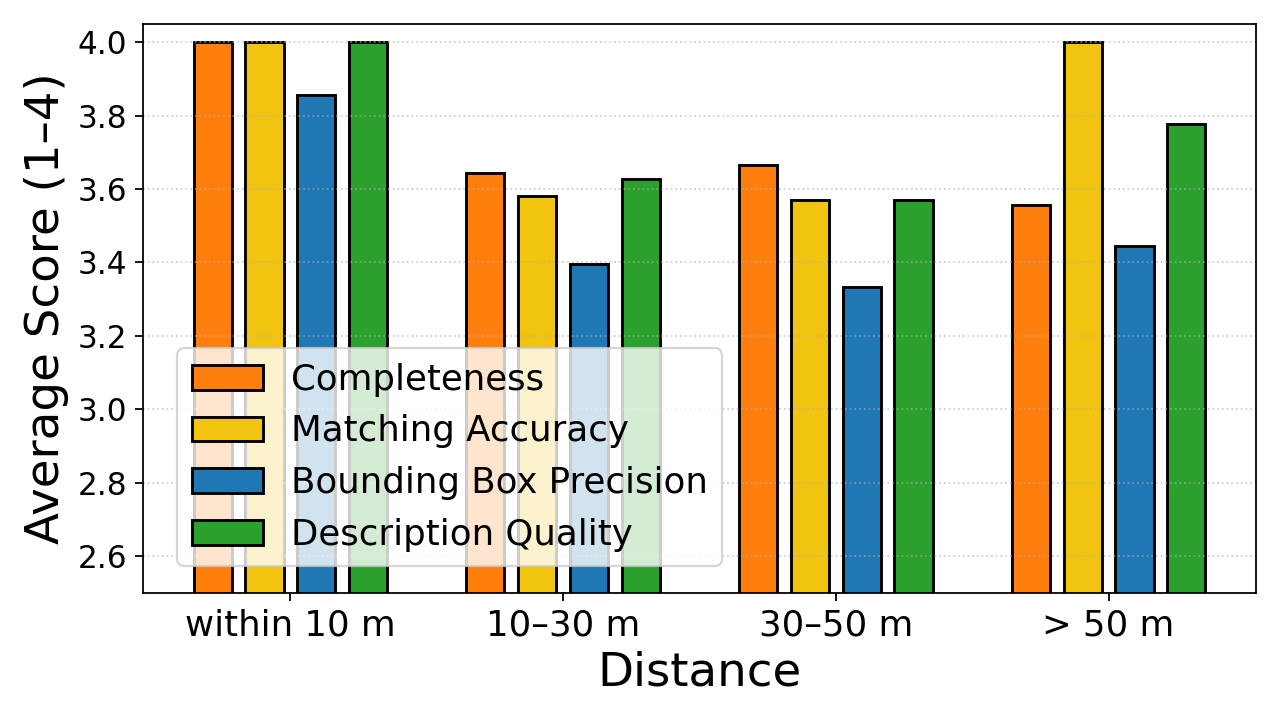}
    \caption{AutoTour performance across different feature distances.}
    \label{fig:distance_curve}
  \end{minipage}
  \hspace{0.2cm}
  \begin{minipage}[b]{0.45\linewidth}
    \centering
    \includegraphics[width=\linewidth]{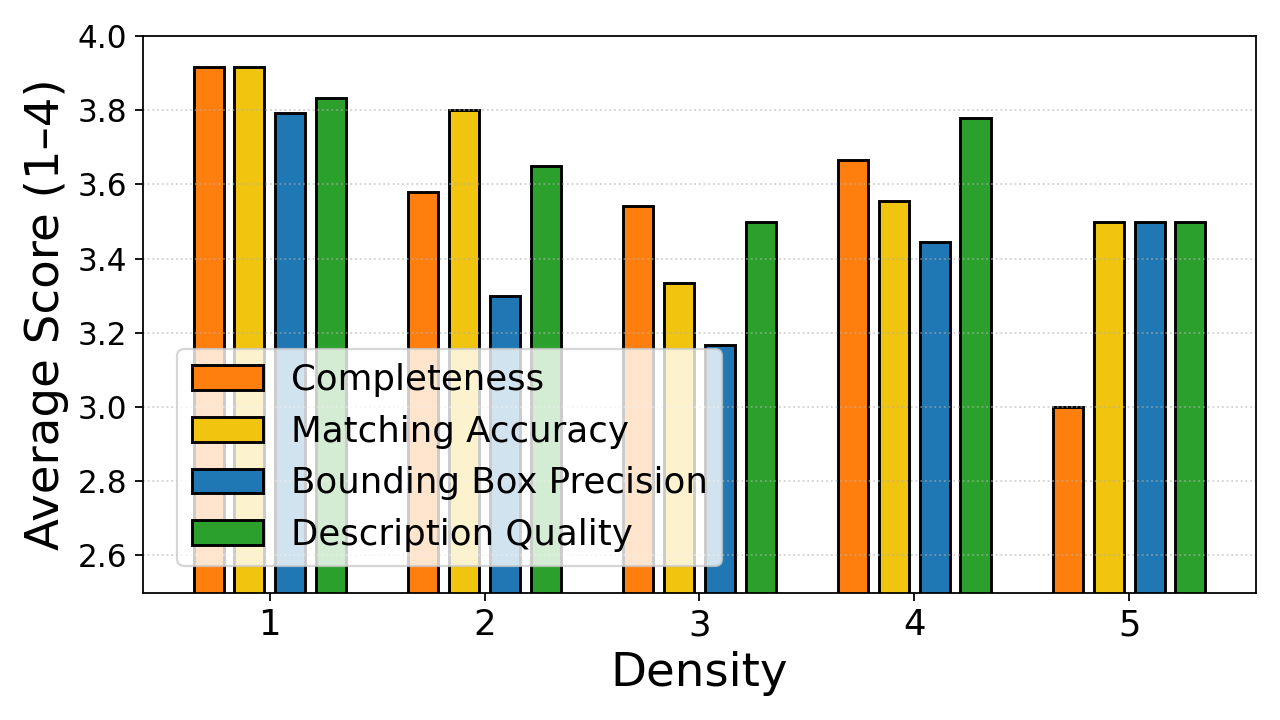}
    \caption{AutoTour performance across different feature densities.}
    \label{fig:feature_density}
  \end{minipage}
  \hfill
\end{figure}
We further examined how the camera–feature distance affects the overall system performance. The evaluation dataset was divided into four distance ranges: within 10 m, 10–30 m, 30–50 m, and greater than 50 m from the photographed subject.
For each range, we computed the average completeness, matching accuracy, bounding box precision, and description quality across all samples.
As illustrated in Fig.~\ref{fig:distance_curve}, AutoTour achieves its best performance for close-range features (within 10 m). While performance slightly decreases for more distant features, the overall scores remain above 3.0 across all ranges, indicating that AutoTour is largely robust to variations in feature distance..

\subsubsection{Impact of feature density}
To evaluate the impact of scene complexity on matching performance, we further analyzed model behavior under varying \textit{feature densities}, defined as the number of identifiable photo features. We considered discrete density levels, specifically 1, 2, 3, 4, and 5 features per photo. Similarly, we adopt the four metrics. As shown in Fig.~\ref{fig:feature_density}, AutoTour maintains consistently high performance, achieving scores above 3.0 in most cases across all metrics. We observe a slight performance degradation as feature density increases, likely due to the higher scene complexity within a single image. A potential solution is to train more specialized models for feature detection and grounding, or alternatively, to guide users to capture photos focusing on a few specific landmarks to achieve optimal results.




\subsection{System Evaluation}
To assess the detailed performance of individual components, we establish a subset of the original dataset with human-annotated ground truth of photo features and their matched OSM feature as ground truths.

\subsubsection{Photo feature detection}.
The experiment compared the performance of three representative models—\textsc{GPT-4o (2024-05-13)}, \textsc{Claude Opus 4 (2025-05-14)}, and \textsc{Grok 3}. Each model was tested under identical settings to ensure fair comparison. We report four key metrics for evaluation: recall, precision, F1-score, and hallucination rate. Recall and precision jointly indicate the accuracy and completeness of visual entity detection, the F1-score provides a balanced measure of overall accuracy, while the hallucination rate reflects the frequency of incorrect or non-existent object descriptions.

Table~\ref{tab:eva:photo} summarizes the quantitative results, and all three models deliver generally competitive results. The Claude~Opus~4 achieves the highest recall, precision, and F1-score, indicating superior completeness and balanced prediction accuracy. GPT‑4o attains the lowest hallucination rate, demonstrating stronger reliability in avoiding spurious object generation, whereas Grok~3 lags in recall compared with the other two models.  Based on these findings, we select Claude~Opus~4 as the default model for our photo feature detection task.
\begin{table}[t!]
\centering
\caption{Performance comparison of three representative Visual-Language Models on the photo feature detection task. Bolded numbers indicate the best scores among the models.}
\label{tab:eva:photo}
\begin{tabular}{c|c|c|c|c}
\toprule
\textbf{Model} &
\textbf{Recall (↑)} &
\textbf{Precision (↑)} &
\textbf{F1-Score (↑)} &
\textbf{Hallucination Rate (↓)} \\ \midrule
\textbf{GPT-4o (2024-05-13)} &
0.864 &
0.830 &
0.850 &
\textbf{0.042} \\ \midrule
\textbf{Claude Opus 4 (2025-05-14)} &
\textbf{0.884}  &
\textbf{0.877} &
\textbf{0.880}  &
0.066 \\ \midrule
\textbf{Grok 3} &
0.747 &
0.872 &
0.800 &
0.053 \\ \bottomrule
\end{tabular}
\end{table}

\begin{table}[t!]
\centering
\caption{Performance comparison between LLMs and the Area‑Overlapping matching method on photo and spatial feature matching algorithm. Bolded numbers indicate the best scores among the models. SD indicates the standard deviation.}
\label{tab:eva:match}
\scalebox{0.85}{
\begin{tabular}{c|c|c|c|c|c|c}
\toprule
\textbf{Metric} &
\textbf{Claude‑Opus‑4} &
\textbf{Gemini‑2.0‑Flash} &
\textbf{GPT‑4o} &
\textbf{LLaMA‑3‑70B} &
\textbf{DeepSeek‑V3} &
\textbf{Area‑Overlapping} \\
\midrule
\textbf{Recall Mean (↑)} &
\textbf{0.816} &
0.721 &
0.712 &
0.712 &
0.696 &
0.776 \\
\textbf{Recall SD (↓)} &
\textbf{0.285} & 0.359 & 0.341 & 0.374 & 0.366 & 0.365 \\ \midrule
\textbf{Precision Mean (↑)} &
0.775 &
0.699 &
0.692 &
0.617 &
0.636 &
\textbf{0.923} \\
\textbf{Precision SD (↓)} &
0.294 & 0.353 & 0.340 & 0.358 & 0.359 & \textbf{0.277} \\ \midrule
\textbf{F1‑score (↑)} &
0.79 &
0.71 &
0.70 &
0.66 &
0.66 &
\textbf{0.84} \\ \midrule
\textbf{Hallucination Rate (↓)} &
\textbf{0.000} &
0.053 &
0.069 &
0.145 &
0.017 &
- \\ \midrule\midrule
\textbf{Token Usage (↓)} &
1380.4 &
1050.2 &
1141.9 &
1200.2 &
1149.8 &
-\\\midrule
\textbf{Latency (↓)} &
10.9s &
18.6s &
5.2s &
13.3s &
10.8s &
\textbf{$\sim$100ms}\\
\bottomrule
\end{tabular}}
\end{table}
\subsubsection{Feature Matching}
To benchmark photo–map feature matching performance, we evaluate two types of solutions: (1) our geometric-based \textit{Area Overlapping Matching} approach, as described in Section~\ref{sec:feature:matching}; and (2) an LLM-based matching scheme, where key feature attributes—such as distance, direction, and textual descriptions—are organized into structured prompts to guide the model in performing correspondence inference. 
matching results are decoded directly from the generated responses. Each method is assessed using four metrics: Recall, Precision, F1-score, and Hallucination Rate. Table~\ref{tab:eva:match} summarizes the quantitative results, including mean and standard deviation values. The comparison covers Claude Opus‑4 (2025‑05‑14), Gemini‑2.0‑Flash, GPT‑4o (2024‑05‑13), LLaMA‑3‑70B, DeepSeek‑V3, and our proposed \textit{Area Overlapping Matching} baseline.  

Overall, the \textit{Area Overlapping Matching} method outperforms all evaluated LLMs across most metrics, demonstrating strong robustness for spatially grounded photo–map alignment. 
Among the LLMs, Claude Opus‑4 achieves the highest recall (0.816) with zero hallucinations, but its precision and consistency remain lower than those of the geometric overlap baseline. 
Other LLMs show moderate-to-low matching accuracy and higher metric variance, exposing limitations in fine-grained spatial reasoning and geometric object correspondence. Importantly, our geometric method introduces no hallucination, executes within one second, and incurs negligible computational cost compared to LLM-based inference, which requires paid model queries. These results indicate that explicit, geometry-aware overlap reasoning can outperform latent multimodal text reasoning in structured spatial alignment tasks.

\subsubsection{Bounding Box Fixing}
In Section~\ref{sec:design:visual}, we propose leveraging VLMs to refine the accuracy of bounding boxes. We evaluated 50 cases and found that in 29.3\% of them, the VLM adjusted the bounding box coordinates. Among these adjusted cases, 55.6\% of cases show improved bounding box accuracy, while 11.1\% result in degraded performance. Overall, incorporating VLMs enhances bounding box accuracy. But it also introduces additional latency, representing a trade-off that is further discussed in Section~\ref{sec:latency}.

\subsection{Overhead}

\subsubsection{Latency}\label{sec:latency}





\begin{figure}[t!]
\centering
\includegraphics[width=1.0\linewidth]{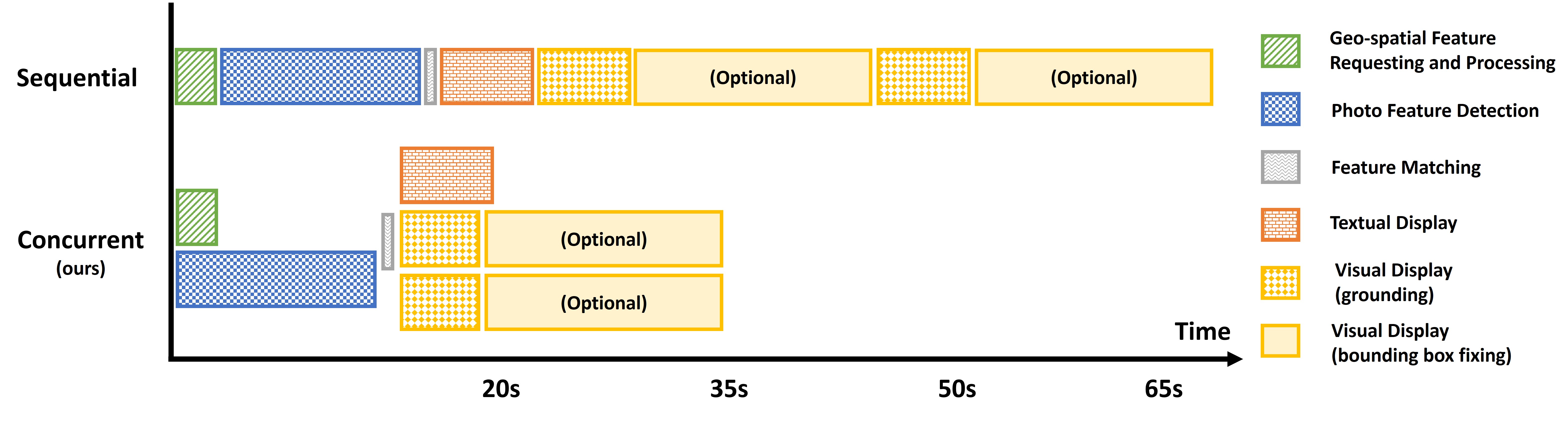}
\caption{Breakdown of AutoTour’s overall latency, showing the time consumption of each processing module.}
\label{fig:time_comparison}
\end{figure}
This section presents the time cost of each module and the overall system latency. To reduce latency, we adopt a \textit{concurrent execution} strategy that enables multiple components to run in parallel when processing multiple features within the same photo. As shown in Fig.~\ref{fig:time_comparison}, while the VLM performs feature identification, the system concurrently executes other tasks such as geographic data retrieval and OSM element pre-processing. Grounding tasks for different features are also processed in parallel. Compared with sequential execution, our concurrent design significantly reduces latency, achieving approximately 20 seconds without bounding box refinement and about 35 seconds when bounding box fixing is applied. To achieve optimal latency, the bounding box fixing component can be optional for some cases. The latency issue is further discussed in the Discussion section.

\subsubsection{Token Usage}
\begin{table}[t!]
\centering
\caption{Estimated token usage and computation cost per photo analysis module.}
\label{tab:token_cost}
\scalebox{0.9}{
\begin{tabular}{c|c|c|c|c|c}
\toprule
\textbf{Module} & \textbf{Model} &
\textbf{\begin{tabular}[c]{@{}c@{}}Prompt\\Tokens\end{tabular}} &
\textbf{\begin{tabular}[c]{@{}c@{}}Completion\\Tokens\end{tabular}} &
\textbf{Price (USD / 1M)} &
\textbf{Est. Cost (USD)} \\
\midrule
Photo Feature Detection & Claude Opus 4 & 1417 & 498 & Input \$15 / Output \$75 & 0.0213 \\
Spatial Feature Extraction & -- & -- & -- & -- & 0.0000 \\ \midrule
Feature Matching & -- & -- & -- & -- & 0.0000 \\ \midrule
Feature Grounding & Qwen-Max & 1725 & 290 & Input \$1.6 / Output \$6.4 & 0.0038 \\
Tour Guide Translation & GPT-4o & 246 & 265 & Input \$2.5 / Output \$10 & 0.0032 \\
\midrule
\textbf{Total} &  &  &  &  & \textbf{0.0283} \\
\bottomrule
\end{tabular}}
\end{table}

Table \ref{tab:token_cost} presents the token usage (including multimodal prompting) and estimated computation cost of each module in the photo analysis pipeline. The largest cost arises from the photo feature detection step using Claude Opus 4, while other modules, such as feature grounding and translation, contribute only marginally. Overall, the total estimated cost per photo analysis is approximately \$0.03, indicating that the system operates with very low computational expense per instance.


\section{Discussion}

\textbf{System latency}. One major limitation of the current prototype lies in its processing latency. The use of large VLMs for feature detection introduces substantial computation time. A potential future direction for improvement is to develop a lightweight, task-specific detection model trained explicitly for buildings and landmarks. Such a model could significantly reduce inference time while maintaining acceptable accuracy, enabling near real-time performance for interactive exploration scenarios.

\textbf{Dependence on GPS accuracy}. The accuracy of \textit{AutoTour} heavily depends on the quality of GPS data. GPS errors can directly affect both distance and direction estimation, thereby influencing feature matching accuracy. This limitation becomes particularly evident in dense urban environments (e.g., downtown areas with high-rise buildings) or indoor settings, where GPS signals are weak or unavailable. Consequently, the system is best suited for outdoor use in open areas where GPS readings are stable. A promising future enhancement would be to calibrate or refine the estimated location using visual cues from photos for position correction.

\section{Conclusion}
In this paper, we presented \textit{AutoTour}, a novel system that enhances user exploration through photo-based landmark annotation and contextual description. By integrating vision-language models with open geospatial data from OpenStreetMap, AutoTour can identify, match, and describe visible landmarks directly on user-captured photos, providing both visual and auditory guidance. Our prototype demonstrates the feasibility of this training-free, multimodal approach for scalable, context-aware travel assistance. AutoTour demonstrates a promising direction for applying LLMs to the physical world by integrating them with real-world sensor data for contextual understanding and interaction.



\bibliographystyle{ACM-Reference-Format}
\bibliography{sample-base}










\end{document}